\documentclass[reprint,amsmath,amssymb,aps,nofootinbib]{revtex4-2}

\usepackage{graphicx}
\usepackage{dcolumn}
\usepackage{bm}
\usepackage{hyperref}
\usepackage{xcolor}
\usepackage{ulem}
\usepackage{tikz}
\usepackage{listings}
\usepackage[font=small,labelfont=bf,justification=justified]{subcaption}
\usepackage[font=small,labelfont=bf,justification=justified]{caption}
\usepackage{multirow}
\usepackage[makeroom]{cancel}
\usepackage{float}

\usepackage{float}
\graphicspath{{Figures/}}

\definecolor{lime}{HTML}{A6CE39}
\DeclareRobustCommand{\orcidicon}{
	\begin{tikzpicture}
	\draw[lime, fill=lime] (0,0) 
	circle [radius=0.16] 
	node[white] {{\fontfamily{qag}\selectfont \tiny ID}};
	\draw[white, fill=white] (-0.0625,0.095) 
	circle [radius=0.007];
	\end{tikzpicture}
	\hspace{-2mm}
}
\foreach \x in {A, ..., Z}{\expandafter\xdef\csname orcid\x\endcsname{\noexpand\href{https://orcid.org/\csname orcidauthor\x\endcsname}{\noexpand\orcidicon}}}

\begin{document}

\preprint{APS/123-QED}

\title{Polytropic fits of modern and unified equations of state}

\author{Lami Suleiman\orcidA{}}
\affiliation{Nicolaus Copernicus Astronomical Center of the Polish Academy of Sciences, ul. Bartycka 18, 00-716 Warszawa, Poland}
\affiliation{Laboratoire Univers et Th\'{e}ories, Observatoire de Paris, Universit\'{e} PSL, Université Paris Cit\'{e}, CNRS, F-92190 Meudon, France}
\def\ls{\textcolor{purple}}

\author{Morgane Fortin\orcidB{}}%
\affiliation{Nicolaus Copernicus Astronomical Center of the Polish Academy of Sciences, ul. Bartycka 18, 00-716 Warszawa, Poland}
\def\mf{\textcolor{blue}}

\author{Constan\c ca Provid\^encia\orcidC{}}
\affiliation{CFisUC, Department of Physics,
University of Coimbra, P-3004 - 516  Coimbra, Portugal}
\def\cp{\textcolor[rgb]{0.0,0.8,0.2}}

\author{Julian Leszek Zdunik\orcidD{}}
\affiliation{Nicolaus Copernicus Astronomical Center of the Polish Academy of Sciences, ul. Bartycka 18, 00-716 Warszawa, Poland}
\def\lz{\textcolor{orange}}
 
\def\everyone{\textcolor{orange}}

\date{\today}

\begin{abstract}
\begin{description}
\item[Background] 
Equations of state for a cold neutron star's interior are presented in three-column tables that relate the baryonic density, the energy density, and the pressure. A few analytical expressions for those tables have been established these past two decades, as a convenient way to present a large number of nuclear models for neutron-star matter. Some of those analytical representations are based on nonunified equations of state, in the sense that the high and the low density part of the star are not computed with the same nuclear model. 
\item[Purpose]
Fits of equations of state based on a piecewise polytropic representation are revised by using unified tables of equations of state, that is to say models which have been calculated consistently for the core and the crust. 
\item[Methods]
A set of 52 unified equations of state is chosen. Each one is divided in seven polytropes via an adaptive segmentation, and two parameters per polytrope are fitted to the tabulated equation of state. The total mass, radius, tidal deformability and moment of inertia of neutron stars are modelled from the fits and compared with the quantities calculated from the original tables to ensure the accuracy of the fits on macroscopic parameters.
\item[Results]
We provide the polytropes parameters for 15 nucleonic relativistic mean-field models, seven hyperonic relativistic mean-field models, five hybrid relativistic mean-field models, 24 nucleonic Skyrme models, and one \textit{ab initio} model. 
\item[Conclusion] 
The fit error on the macroscopic parameters of neutron stars is small and well within the estimated measurement accuracy from current and next generation telescopes.
\end{description}
\end{abstract}

\maketitle


\section{Introduction}
The equation of state (EoS) plays a key role in modeling a neutron star's macroscopic parameters. Despite continuous efforts to push the limits of nuclear experiments, conditions of density and temperature in the deepest layers of neutron stars remain out of reach for laboratories. Multimessenger astronomy provides a chance to probe deep inside these extremely compact stars: the relativistic hydrodynamics equations operate as a bridge between the unknown microphysics of high-density neutron-rich matter, and observable macroscopic parameters. The measurement of the mass $M$, the radius $R$, the tidal deformability $\Lambda$, and the moment of inertia $I$ paired with hydrodynamics equations in gravity theories, have turned neutron stars (NSs) into extra-terrestrial laboratories for high density nuclear physics. 

Neutron stars are observed in various wavelengths of the luminosity spectrum. On one hand, they are observed in the x-ray spectrum. The spatial telescope X-ray Multi Mirror-Newton (XMM-Newton \citep{XMMNewton}) from the European Spatial Agency, and its American counterpart Chandra \citep{Chandra}, have been operating for more than 20 years, providing data for isolated, accreting, and highly magnetized neutron stars. The younger Nuclear Spectroscopic Telescope Array (NuSTAR \citep{NUstar}), and Neutron-star Interior Composition ExploRer (NICER \citep{NICER}) observe respectively in hard and soft x-rays. The next generation of x-ray telescopes is already on its way: the Enhanced X-ray Timing and Polarimetry (eXTP  \citep{eXTP}) mission is scheduled to be launched in 2027, and the highly anticipated Advanced Telescope for High ENergy Astrophysics (ATHENA \citep{Hauf2012}) with a state-of-the-art X-ray Integral Field Unit (X-IFU) spectrometer will provide unprecedented spectral resolution for a wider effective area, hopefully in the 2030s. On the other hand, neutron stars are observed as radio sources, for example, in the observatories of Parkes, Green Bank, and Nancay. After almost 60 years of operation in Porto Rico, Arecibo fell in November 2020; fortunately, the largest worldwide telescope, the Square Kilometer Array (SKA  \citep{SKA}) will be operational in a few years and will include the 2017 launched Chinese contribution Five-hundred-meter Aperture Spherical radio Telescope (FAST  \citep{FAST}). Gravitational wave (GW) detection is the most recently explored area for compact object messaging: the LIGO Scientific Collaboration, the Virgo Collaboration, and the KAGRA Collaboration (LVK \citep{LVK}) have provided promising results with the detection of double-neutron-star binary mergers, and will keep on with run O4 starting in early 2023, and then O5. The Einstein telescope \cite{ET2020}, and its American counterpart the Cosmic Explorer \citep{CE2021}, are both scheduled to start observing in the mid 2030s, and should detect continuous gravitational waves.

\begin{figure}
    \resizebox{\hsize}{!}{\includegraphics{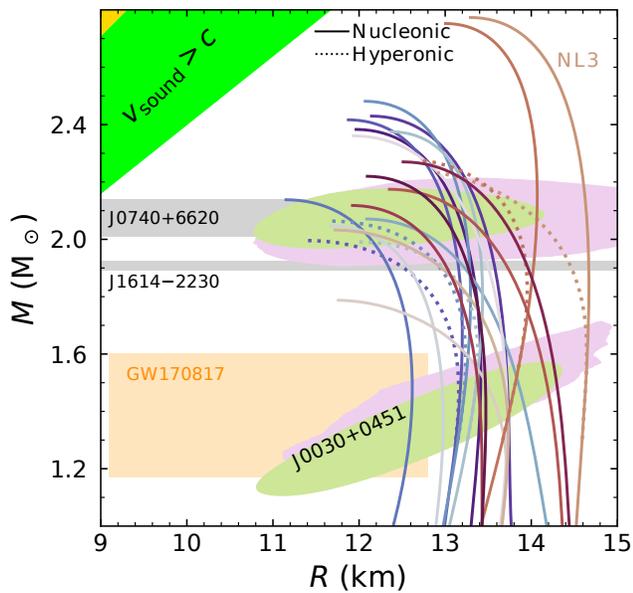}}
    \caption{Mass-radius (MR) sequence for the unified nucleonic and hyperonic relativistic mean-field EoSs used in this paper. Mass measurements of J0740$+$6620 and J1614$-$2230 within 1$\sigma$ precision are presented, as well as NICER mass-radius contours of J0740$+$6620 and J0030$+$0451 within 1$\sigma$ (see text for details), and the mass-radius constraint on GW170817. The exclusion regions for finite pressure and subluminal EoSs are represented respectively in yellow and green.}
    \label{fig:MRrmf}
\end{figure}

Part of the outer crust of neutron stars is well constrained by laboratory measurements of mildly neutron rich nuclei. However, the inner crust and the core are respectively poorly, and very poorly constrained which leads to many different supra-nuclear theories: nucleonic matter, confined or deconfined quarks, hyperonic matter, strange quark matter, etc. There exists dozens of equations of state for cold and highly dense neutron-star matter, a large number of them are gathered in an online data base named \href{https://compose.obspm.fr/}{CompOSE} \citep{typel2022compose}. The crust equation of state is more difficult to compute than the core's because it requires a treatment of inhomogeneous matter. For that reason, one can find nonunified constructions of equations of state: the core and the crust are not computed using the same nuclear model; the common practice is to attach an already established crust to the core equation of state. Nonunified constructions are oftentimes found in the literature, but have been shown to result in large errors in the modeling of macroscopic parameters, see Ref.~\cite{Suleiman2021}. 

The output format for computations of cold matter equations of state is a three column table with the baryonic density $n$, the energy density $\epsilon$, and the pressure $P$. However, an analytical form of tabulated equations of state is convenient, particularly for neutron-star simulations. To establish analytical representations of equations of state, one chooses a parametrized expression, and then adjusts parameters to the tabulated equation of state. Having one expression with easily comparable parameters is also a practical way to compare microscopic and macroscopic features of neutron stars. Adjusting the parametrized representation is called a fit: a few have been proposed so far, e.g., spectral fits as presented in Ref.~\cite{Lindblom_2018}, and piecewise polytropic fits by Ref.~\cite{Read09}. 

In this paper, piecewise polytropic fits based on nonunified constructions are revised, and new fits for 52 unified equations of state of neutron star's matter are presented. In Sec.~\ref{Choice}, the most relevant microscopic and macroscopic features required for a realistic equation of state are presented. After a brief overview of the various frameworks for nuclear interaction modeling used to compute equations of state, we present the equations of state used in this paper. Piecewise polytropic fits as well as the relation between polytropic parameters, and quantities $n$, $\epsilon$ and $P$, are presented in Sec.~\ref{PP}. The importance of using unified equations of state, on the accuracy of macroscopic parameters is investigated; this section ends with details on our fitting method. In Section.~\ref{macro}, the accuracy of fits on the mass, the radius, the moment of inertia, and the dimensionless tidal deformability calculated within the framework of Einstein's theory of general relativity is presented for key astrophysical quantities. The role of nonunified constructions on the accuracy of so-called "universal" relations is discussed. The parameters for the fits are presented in tables in Appendix \ref{append}.

\section{Choice of equations of state} \label{Choice}
The three basic physics rules that an EoS is required to meet are: 
\begin{itemize}
    \item thermodynamic consistency: the first law of thermodynamics must be fulfilled,
    \item causality: interpreted as subluminal EoS (${v_{\rm sound}<c}$), for details concerning Lorentz invariance and causality see Ref.~\cite{Haensel2007}, 
    \item Le Chatelier's principle which states that the energy increases with pressure.  
\end{itemize}
The natural limit up to which one can study the properties of dense matter using NS observations is set by the properties of matter at the center of the star with maximum mass configuration - the corresponding value of the central baryon number density is denoted $n_{max}$.

Astrophysical and nuclear laboratory data can impose more constraints to ensure that the equation of state is realistic for NS matter.


\begin{figure}
    \resizebox{\hsize}{!}{\includegraphics{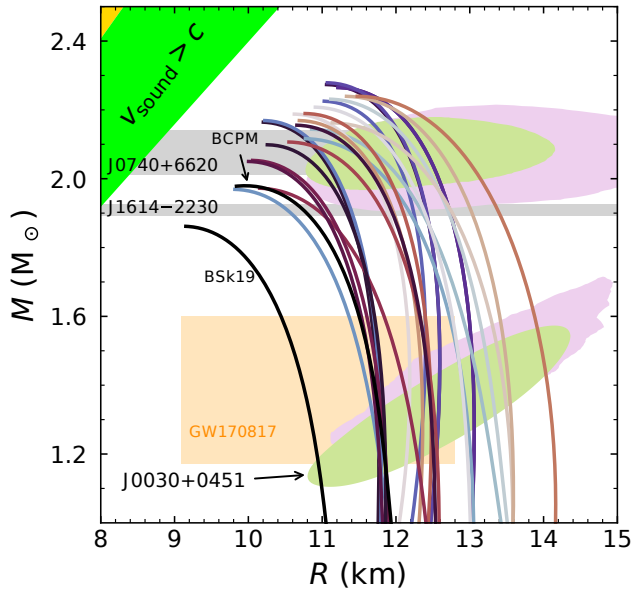}}
    \caption{Mass-radius sequence for unified Skyrme EoSs and the \textit{ab-initio} model BCPM. Mass measurement of J0740$+$6620 and J1614$-$2230 within 1$\sigma$ precision are presented, as well as NICER mass-radius contours of J0740$+$6620 and J0030$+$0451 within 1$\sigma$ (see text for details), and mass-radius constraint on GW170817. Exclusion regions for finite pressure and subluminal EoSs are represented respectively in yellow and green.}
    \label{fig:MRSkyrme}
\end{figure}

\subsection{What makes a reasonable neutron-star equation of state from the macroscopic point of view ?}

The macroscopic parameters of cold NSs are computed by injecting the EoS into the equations of hydrodynamics equilibrium within a given theory of gravity. Here we adopt Einstein's theory of general relativity. Comparing modelled macroscopic parameters to their measurement provides information on the NS matter EoS. More to the point, the quality of an EoS relies on its ability to be consistent with observations of the NS's astrophysical quantities. 

\begin{figure}
    \resizebox{\hsize}{!}{\includegraphics{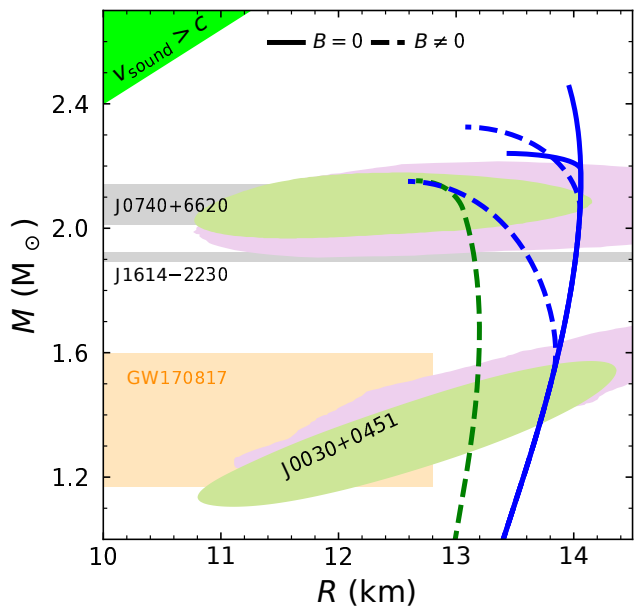}}
    \caption{MR sequence for the unified hybrid relativistic mean-field EoSs used in this paper. Mass measurements of J0740$+$6620 and J1614$-$2230 within 1$\sigma$ precision are presented, as well as NICER mass-radius contours of J0740$+$6620 and J0030$+$0451 within 1$\sigma$ (for details see text), and the mass-radius constraint on GW170817. NL3 models are in blue and DD2 models are in green. The pressure bag constant $B$ is either set to zero (plain lines) or takes a nonzero value (dotted lines). The exclusion region for subluminal EoSs is represented in green.}
    \label{fig:MRhybrid}
\end{figure}

The modeling of a NS's macroscopic parameters treated within general relativity imposes a first constraint on the mass/radius ratio: it must always be larger than that of Schwarzschild which excludes any mass/radius relation for which ${R < 2 GM/c^2}$. A second constraint is imposed by the finite nature of the pressure, and renders ${R> 9GM/4c^2}$ (yellow region in Figs.~\ref{fig:MRrmf},\ref{fig:MRSkyrme}). A third constraint on the mass/radius ratio is imposed to ensure a subluminal EoS: ${R>2.9 GM/c^2}$ (green region in Figs.~\ref{fig:MRrmf},\ref{fig:MRSkyrme},\ref{fig:MRhybrid}).

An important test for an EoS is to reach the highest NS mass measured: in this article, we use the massive millisecond pulsar J1614$-$2230 \citep{Arzoumanian_2018} measured with a mass of $1.908\pm 0.016$\;M$_{\odot}$ (solar mass M$_{\odot}$) as a selection criterion. Sources with larger masses have been reported, such as the millisecond pulsar J0740$+$6620, whose mass was previously measured with relativistic Shapiro delay at $2.14_{-0.09}^{+0.10}$\;M$_{\odot}$ \citep{Cromartie_2019}, but was recently revised to a $2.08 \pm 0.07$\;M$_{\odot}$ \citep{fonseca2021refined}. The source J0348$+$0432 was reported with a mass of $2.01\pm 0.04$\;M$_{\odot}$ \citep{Antoniadis2013} and J1810$+$1744 has been measured with a mass of $2.13\pm0.04$\;M$_{\odot}$ \citep{Romani_2021}; their mass measurement technique is based on a highly model dependent analysis of the companion's photometry (white dwarf). We refer to Fig.~1 of Ref.~\cite{Suleiman2021} for an overview of NS measured masses. The maximum mass of a cold nonrotating star was constrained by gravitational wave (GW) detection \citep{Abbott_2017} of the double NS binary GW170817 in Ref.~\cite{Khadkikar_2021}, which highlighted the importance of finite temperature in the relation between the gravitational mass Keplerian limit and the maximum mass of a nonrotating cold star. 

The radius is directly connected to the EoS's stiffness. For a significant range of pressures typical of neutron-star interior, the corresponding densities are smaller for a stiff EoS than for a soft one. As a result, from the Tolman-Oppenheimer-Volkoff equations, the thickness of this region is larger for stiff EoSs which can lead to a larger radius of the NS. A strong correlation exists between the pressure and the radius at densities $[1-2.5]n_0$ (saturation density $n_0=0.16$\;fm$^{-3}$), as shown by Ref.~\cite{Lattimer2001}. The relations between the total mass and the total radius for the EoSs used in this paper are presented in Figs.~\ref{fig:MRrmf}, \ref{fig:MRSkyrme} and \ref{fig:MRhybrid}. The radius and mass of two sources -PSR J0030$+$0451 and PSR J0740$+$6620, have been reported by the NICER telescope by two teams each. The measurement technique is based on an analysis of the surface emission of the pulsar, precisely of its hot spots. The source J0030$+$0451 was reported by Ref.~\cite{Riley_2019} to have a mass of  ${1.34^{+0.15}_{-0.16}\;{\rm M}_{\odot}}$ and radius of ${12.71^{+1.14}_{-1.19}\;{\rm km}}$ and reported by Ref.~\cite{Miller_2019} to have a mass of ${1.44_{-0.14}^{+0.15}\;{\rm M}_{\odot}}$ and a radius of ${13.02^{+1.24}_{-1.06}\;{\rm km}}$  within 1$\sigma$ precision. The source J0740$+$6620 was reported by \cite{riley2021nicer} to have a radius of ${13.7^{+2.6}_{-1.5}\;{\rm km}}$ and by \cite{miller2021radius} to have a radius of ${12.39^{+1.3}_{-0.98}\;{\rm km}}$ within 1$\sigma$ precision; prior knowledge from XMM-Newton telescope on the mass for this source was used. Contours for those sources are presented in Figs.~\ref{fig:MRrmf}, \ref{fig:MRSkyrme} and \ref{fig:MRhybrid}; for the source reported by Ref.~\citep{Miller_2019}, the contours for two signal analysis are given because no preference for one or the other was significant.  Because the uncertainty for the radius is quite large, measurements serve more as a proof of concept for an elegant radius determination, than a conclusive constraint on NS matter. In Refs.~\cite{Abbott_2018, Abbott_2020}, an indirect estimation of the radius was established from GW170817 gravitational wave (GW) detection; the authors either used "universal" relations established by Ref.~\cite{Yagi17} to present a radius with 3.5\;km error bars, or a collection of EoSs in a Bayesian analysis to give a likelihood for the radius. As a substitute for radius measurement, a series of papers attempt to impose limits on the radius. Reference \cite{Steiner_2013} used a  prior distribution of EoSs to obtain a radius interval of $[10.4-12.9]$\;km for a $1.4$\;M$_{\odot}$ NS with a 95\% confidence level. A similar approach is used by Ref.~\cite{Guillot_2013} in a Monte Carlo analysis with five low mass x-ray binary sources, to extract a minimum radius of 9.1$_{-1.5}^{+1.3}$\;km within a 90\% confidence level. Unfortunately, those limits are established from a small set of sources. Reference \cite{Haensel2009} established a constraint on the radius, based on the assumption that NS's rotation follows a Keplerian frequency; the authors conclude that R$_{1.4} \leq 9.3$\;km for the radius of a 1.4\;M$_{\odot}$. However, NSs might not follow such frequencies when the rotation of the star is associated with GW emission or when triaxial deformability sets on.

The tidal deformability $\Lambda$ is the propensity of a star to be deformed by a neighboring gravitational field. For the binary NS merger GW170817, the tidal deformabilities of the stars were extracted from the inspiral waveform. This detection has indicated a preference towards a soft EoS, see \cite{Malik18,Raithel18}. Conciliating a high enough maximum mass with a soft EoS, while keeping a small radius is a delicate question of balance for the inner core composition. Constraints on the tidal deformability of a 1.4\;M$_{\odot}$ NS are discussed in Ref.~\cite{Abbott_2018}. 

The moment of inertia $I$ is a parameter that has not been measured so far. It requires a monitoring of the relativistic features of the binary orbit (the more compact the better) over a long period of time, see \cite{Greif_2020}. This parameter would be best measured in a pulsar binary such as the famous PSR J0737$-$3039 \citep{Kramer2021}. One can extract $I$ using "universal" relations: \cite{Silva2021} have used a Markov chain Monte Carlo analysis of NICER J0030$+$0451 data to establish a radius distribution for the star of mass 1.3381\;M$_{\odot}$ in double binary PSR J0737$-$3039; from the compactness $C=GM/Rc^2$, they estimate the moment of inertia.

\subsection{What makes a reasonable NS EoS from the microscopic point of view ?}

A large portion of the outer crust is constrained by laboratory measurements of mildly rich nuclei; for example,  Ref.~\cite{ame2020} has catalogued such results in a table with thousands of nuclei data. All models for NS interior should therefore be similar for the first few layers of the crust.

The symmetry energy and the slope of the symmetry energy at saturation density, respectively $J$ and $L$, are two microscopic parameters used to describe an EoS. The symmetry energy is defined as the difference between the energy per baryon calculated for pure neutron matter, and the energy per baryon calculated for symmetric matter. A series of laboratory experiments can constrain both those parameters; we refer to Table~II and Fig.~8 of Ref.~\cite{Oertel2017} for a systematic approach. We refer to Ref.~\cite{Tsang2012} for a compilation of constraints on the symmetry energy obtained from experiments and theory. In the following, we briefly present, nonexhaustively, how both these parameters can be constrained by laboratory data.

The binding energy described by the Finite Range Droplet Model (FRDM) includes symmetry-related terms whose values can be explored by using large tables of nuclei data. The formula for the FRDM binding energy presented in Ref.~\cite{Moller2012} is paired with the table of nuclei of Ref.~\cite{ame2012} to extract: $L=70\pm15$MeV and $J=32.5 \pm 0.5$MeV. Coulomb effects are linked to the surface symmetry term in the FRDM (see Ref.~\cite{Danielewicz2003}, Sec.~IIC); experiments are performed on isobaric nuclei \cite{Danielewicz2014} in order to alleviate this entanglement. The same method has been used with Skyrme forces in Ref.~\cite{Kortelainen2010}.

The Heavy Ion Collision (HIC) of nuclei such as gold has introduced constraints on symmetric matter over the saturation density. The collision of nuclei such as isotopes of tin allows one to probe the asymmetry between the number of protons and the number of neutrons \cite{Tsang2009}. Recently, the spectral pion ratio of tin isotopes has been used to determine the slope $L$ of the symmetry energy at saturation density  and, at a 95\% confidence level, $42 < L < 117$ MeV, an interval consistent with the conclusions drawn in Ref.~\cite{Tsang2012}.

\begin{figure}
\resizebox{\hsize}{!}{\includegraphics[scale=0.3]{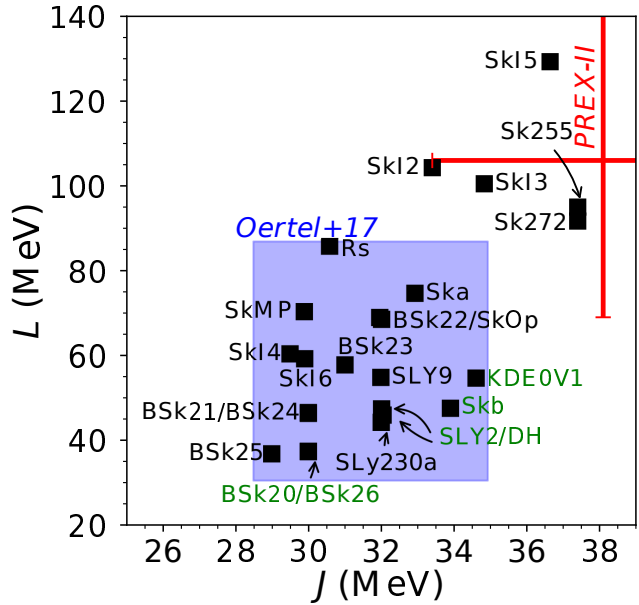}}
\caption{Symmetry energy and its slope at saturation density respectively denoted $J$ and $L$ for Skyrme models used in this paper. Experimental data constraints are presented: in blue is the compiled constraint presented in Ref.~\cite{Oertel2017}, and in red that of PREX-II. The names of the EoSs in green refer to nucleonic models that do not permit the Direct Urca process. }
\label{fig:JLSkyrme}
\end{figure}

\begin{figure}
\resizebox{\hsize}{!}{\includegraphics[scale=0.3]{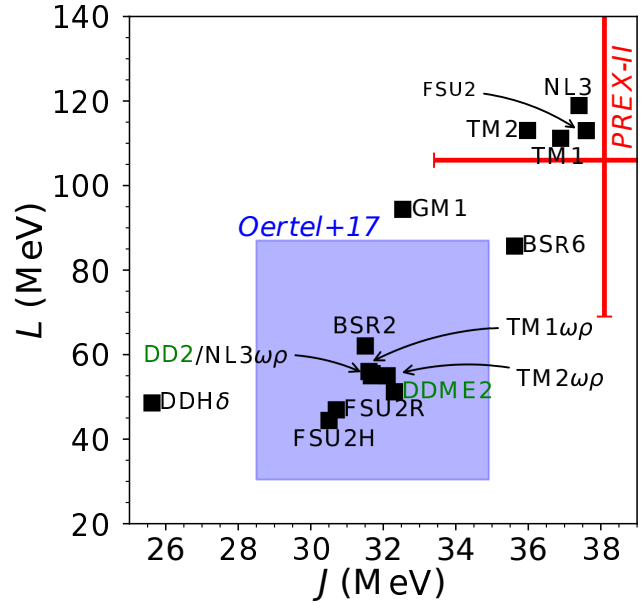}}
\caption{Symmetry energy and its slope at saturation density respectively denoted $J$ and $L$  for relativistic mean-field models used in this paper. Experimental data constraints are presented: in blue is the compiled constraint presented in \cite{Oertel2017}, and in red that of PREX-II. EoS's name in green refer to nucleonic models that do not permit the Direct Urca process. }
\label{fig:JLrmf}
\end{figure}

Neutron-rich nuclei are particularly interesting to investigate the symmetry energy when they present an asymmetric number of neutrons and protons, as is the case of Sn isotopes or $^{208}$Pb that closes its nucleon shells, which simplifies the nucleus structure. An asymmetry in favor of neutrons implies that the nucleus will present a large difference in the radius distribution of neutrons and protons, also called neutron skin. There are a few different ways of measuring the neutron skin thickness, one of which is to see how electroweak parity of $^{208}$Pb is violated by polarized electrons in the experiments PREX-I and PREX-II \cite{PREX1, PREX2} and of $^{48}$Ca in the experiment CREX \cite{crex}. To extract $L$ from neutron skin thickness measurements, the correlation between the two quantities is exploited via a fit established within a nuclear theoretical framework. In Ref.~\cite{Chen2010}, the Skyrme Hartree-Fock model is used on measurements of Sn isotopes to constrain the relation between $J$ and $L$. In Ref.~\cite{Reed_2021}, the FSU2Gold relativistic mean-field parametrization is used on PREX-II data to extract $J=38.1\pm4.7$\;MeV and $L=106\pm37$\;MeV; this result is in tension with other nuclear experiment constraints. An analysis of the compatibility between PREX-I and PREX-II and CREX experiments and other experiments determining $J$ and $L$ is discussed in Ref.~\cite{crex2}. The values of $J$ and $L$ for Skyrme and relativistic mean-field models used in this paper are presented respectively in Figs.~\ref{fig:JLSkyrme} and \ref{fig:JLrmf}. The laboratory constraints from Ref.~\cite{Oertel2017}, and PREX-II data are shown.

Constraints on $L$ and $J$ can be combined with NS luminosity observations to explore the core composition. Indeed, Direct Urca (DUrca) which is a neutrino-emissive rapidly cooling process, is required to explain the cooling of accreting NSs \cite{Fortin_2018}. This process is permitted if the proton fraction is high enough, and is therefore triggered at a given value of the density $n_{\rm DUrca}$. This threshold is constrained by the density dependence of the symmetry energy, such that a large $L$ favors a large proton fraction and a process allowed for lower $n_{\rm DUrca}$ -equivalently NS mass. The presence of hyperons in the core implies that there is no need for an elevated $L$ to trigger the DUrca process, it necessarily appears for a $n_{\rm DUrca} < n_{\rm max}$, for details see e.g., Ref.~\cite{Fortin2021}. In the set of EoSs presented in this paper, nucleonic models DD2, DDME2, BSk20, BSk26, KDE0v1, SLy2, DH and Skb do not permit DUrca ($n_{\rm DUrca} > n_{\rm max}$); they are presented in green in Fig.~\ref{fig:JLrmf} and \ref{fig:JLSkyrme}.

The collective motion of nuclei is a source of giant resonances: let there be an exterior isoscalar monopole operator, the strength function of excited states in response to that operator is directly linked to the nuclear incompressibility $K$ for which experimental data are available, (see Tables I and II of Ref.~\cite{Garg_2018}); again, the relation between experimental data and $K$ is established within a theoretical framework (e.g., Skyrme or Gogny forces). Constraints on $L$ and $J$ can also be extracted, such as presented in Refs.~ \cite{Drischler2020,Trippa2008}.

There are also some attempts at including results from cold atom experiments to constrain a part of the low density EoS using the unitary Fermi gas approach. The idea is to consider that low-density neutron matter can be solely characterized by an infinite range scattering and can be considered as a unitary Fermi gas -see Chap.~2 of Ref.~\cite{Gandolfi_2015} for details- in which the energy per nucleon is determined by a single and universal parameter. There are, however, no considerations of lattice nor clusters with this approach, which are essential in the understanding of crust physics.

Techniques to infer knowledge on microscopic parameters of EoS from astrophysical measurements were recently designed. Bayesian inference statistical analysis of GW signal data helps provide constraints on the EoS; we refer to Ref.~\cite{Golomb2022} for a recent study of this technique, and to Ref.~\cite{Vitale2020} and reference therein for a review. GW170817 detection was used in Ref.~\cite{Raithel_2019} to constrain $L$: the effective tidal deformability of the binary is strongly correlated to that quantity and GW170817 indicates a preference for a low $L$. Reference \cite{Malik18} discovered strong correlations between the tidal deformability and linear combination of pairs of nuclear parameters of different orders. References \cite{miller2021radius, riley2021nicer} related to NICER observation of J0740$+$6620, used a Bayesian analysis based on likelihood of measurement to infer high density EoS properties. Perfect knowledge of the EoS under half the saturation density ($n_0/2 = 0.08$ fm$^{-3}$) is assumed; over $n_0/2$, a parametrized EoS and Gaussian-process-based models are used. A similar technique is used by Ref.~\cite{essick2021} with Chiral Effective Theory (CET) constraints. Note that these techniques depend strongly on the priors chosen.

\subsection{The variety of core composition} 
In the innermost parts of NSs, densities can go as high as 15$n_0$. Such a supranuclear framework is out of reach for laboratories which leaves the description of NS cores open to various composition hypotheses. In the present paper, we use the following three categories: nucleonic, hyperonic, and hybrid models for the core composition. 

In addition to nucleons in the core, the presence of hyperons -baryons with at least one strange quark- softens the core EoS, which induces a smaller radius; hyperons in NSs were first introduced in the 1960s. Because the nucleonic Fermi pressure is higher than the hyperonic pressure at fixed density, allowing the presence of hyperons results in a smaller radius. Softening the EoS, however, leads to a lower maximum mass. This is a problem referred to as the "hyperon puzzle", which can be counteracted if one finds a way to stimulate hyperonic pressure. One way to do so is to instigate repulsion from the nature of baryon interactions, see Refs.~\cite{Chatterjee_2016,Vidana_2013}. Hyperons are expected to appear over 2$n_0$ through a series of reactions involving nucleons such as (but not exclusively)
\begin{align} \label{eq:hyper}
    &p + e^- \to \Lambda + \nu_e  \;,  \nonumber \\
    &p + e^- \to \Sigma^0 + \nu_e  \;, \\
    &n + e^- \to \Sigma^- + \nu_e \;. \nonumber
\end{align}
Those reactions have their inverse thus creating loops of high-neutrino-emissive processes; the role of hyperonic neutrino emission on the cooling of NSs was studied by Ref.~\cite{Raduta_2017}. Hyperons $\Lambda$(uds), $\Sigma^0$(uds), and $\Sigma^-$(dds) once created in Eqs.~\eqref{eq:hyper}, can themselves be sources of double strange $\Xi^-$(dss) hyperons. Laboratory measurements of hyperons are performed in heavy-ion collision experiments at the Thomas Jefferson National Accelerator facility (Jlab, USA), the Mainz Microtron accelerator (MAMI-C, Germany), and the Japan Proton Accelerator Research Complex (J-Parc, Japan); for details see Ref.~\cite{Vidana2021}. However, only very short lived hypernuclei can be measured, making it difficult to give solid constraints on the parametrization of YN (hyperon-nucleon) and YY (hyperon-hyperon) interactions. For example, no scattering has been measured, which is needed to accurately calculate hyperon properties. Attempts at extracting the strange composition of NSs from observations of mass, radius and thermal evolution have concluded that the radius of the star decreases linearly with the increase of the total hyperon content, see Ref.~\cite{Fortin2020}.

In the inner core, a phase transition from confined (baryons) to deconfined quarks can be included: the core is therefore made of (strange-)quark matter and EoSs are called hybrid. 

A common approach to compute hybrid EoSs is the relativistic mean-field (RMF) approach. Quark matter is frequently described  in the framework of the  Nambu-Jona-Lasinio (NJL) model, the quark-meson model or the MIT bag model. The hybrid EoSs we consider in the present study were built in Refs.~\cite{Pereira2016,Ferreira:2021osk} within the SU(3) NJL model for quark matter and a RMF model for hadronic matter.  The confined phase follows the same Lagrangian density as nucleonic or hyperonic (strange quark) matter. The Lagrangian density of deconfined quarks in the NJL model includes four-quark scalar and pseudoscalar interaction terms with coupling constant $G_S$, four-quark vector and pseudovector interaction terms -both vector-isoscalar (VP) with coupling constant $G_\omega$ and vector-isovector (VIPI) with coupling constant $G_\rho$ are considered-,  and the six-quark t'Hooft term that ensures that the  axial symmetry U(1)$_A$ is broken. In the NJL models, the pressure  and energy density are defined up to a constant B. The pressure bag constant B is chosen to ensure either that the effective pressure falls to zero when the baryon chemical potential vanishes (B0) or to impose another type of constraint such as fixing the deconfinement baryonic density (B). Finally, the ratio between the vector and scalar coupling constants, in particular, $\xi=G_\omega/G_s$ and $\eta=G_\rho/G_s$, are parameters that characterize the models and define the intensity of the VP and VIPI channels, respectively (for more details, see Refs.~\cite{Pereira2016} and \cite{Ferreira:2020evu}). In the following, we designate the hybrid EOS by the hadronic EoS-B$x-100\xi-100\eta$, where $x$ characterizes the magnitude of the bag constant in MeV/fm$^{3}$. The quark phase transition induces a density jump that ensures a softening of the EoS. If this jump exceeds a critical value, the softening is such that the MR sequence presents a branch which is partly unstable (with respect to radial oscillations). In this case, a single EoS corresponds to the two stable branches of stellar models, and twin stars with the same mass but different radius can exist.

Hybrid stars have also been described within {\it ab-initio} approaches such as the Brueckner-Hartree-Fock many-body theory with realistic two-body and three-body forces \cite{Maieron:2004af,Chen:2011my} (for other {\it ab-initio} approaches see  the reviews \cite{Oertel2017,Burgio:2021vgk}).

\subsection{Set of equations of state}\label{sec:allEOS}

The large number of available EoSs stems from two variables: the core composition, and the nuclear theory used for computations. There are two approaches to compute particle interaction of NS matter. The microscopic one is based on many-body interactions: such models are time-consuming and only accurately apply to homogeneous matter. The phenomenological approach consists of theories based on parameters adjusted to reproduce the properties of matter measured in laboratories. 

The RMF description of hadronic matter is a phenomenological approach in which particles are considered to be immersed in a common potential established in quantum field theory. The Lagrangian density for nucleonic matter contains terms for (naked) baryons, leptons and quarks, as well as terms for interacting mesons of type $\sigma$ (scalar), $\omega$ (vector-isoscalar), and $\rho$ (vector-isovector), with self and cross interactions, see Ref.~\cite{Fortin2017}.

The EoS is obtained from the Lagrangian density
\begin{equation}
    {\cal L}=\sum_{i=p,n} {\cal L}_i + {\cal L}_{\sigma} + {\cal L}_{\omega}  + {\cal L}_{\rho} + {\cal L}_{\sigma\omega\rho}+{\cal L}_{lep} \, ,
\end{equation}
where the first term ${\cal L}_i$ is the nucleonic term, the last term ${\cal L}_{lep}$ is the leptonic contribution, and the middle terms refer to the mesonic contributions. The nucleon Lagrangian density reads
\begin{equation}
    {\cal L}_i=\bar \psi_i\left[\gamma_\mu i D^{\mu}-M^*\right]\psi_i \, ,
\end{equation}
with the covariant derivative ${i D^{\mu}=i\partial^{\mu}-g_\omega \omega^{\mu}- \frac{g_{\rho}}{2}  {\boldsymbol\tau} \cdot \mathbf{\rho}^\mu \, }$, ${g_\sigma}$, $g_\omega$ and $g_\rho$ being the meson-nucleon couplings. $\boldsymbol \tau$  the SU(2) isospin matrices, and the effective mass is ${M^*=M-g_\sigma \sigma  \, ,}$ with $M$ being the vacuum nucleon mass. The leptonic Lagrangian density is given by
\begin{equation}
{\cal L}_{lep}=\sum_{i=e,\mu}\bar \psi_i\left[\gamma_\mu i \partial^{\mu}-m_i\right]\psi_i \, ,
\end{equation}
where the sum is over electrons and muons, and  $m_i$ is their mass. The mesonic contributions  are:
\begin{eqnarray}
{\cal L}_\sigma& = &+\frac{1}{2}\left(\partial_{\mu}\phi\partial^{\mu}\sigma -m_\sigma^2 \sigma^2 - \frac{1}{3}\kappa \sigma^3 -\frac{1}{12}\lambda\sigma^4\right), \nonumber\\
{\cal L}_\omega& = &-\frac{1}{4}\Omega_{\mu\nu}\Omega^{\mu\nu}+\frac{1}{2} m_\omega^2 \omega_{\mu}\omega^{\mu} + \frac{\zeta}{4!}\zeta g_\omega^4 (\omega_{\mu}\omega^{\mu})^2, \nonumber \\
{\cal L}_\rho & = &-\frac{1}{4}\mathbf B_{\mu\nu}\cdot\mathbf B^{\mu\nu}+\frac{1}{2} m_\rho^2 \mathbf \rho_{\mu}\cdot \mathbf \rho^{\mu}+\frac{\xi}{4!} g_\rho^4 (\mathbf\rho_{\mu}\rho^{\mu})^2, \nonumber\\
{\cal L}_{\sigma\omega\rho} & = &  \Lambda_\omega g_{\rho}^2 g_{\omega}^2 \omega_\mu\omega^\mu \boldsymbol{\rho}^{\,\mu} \cdot \boldsymbol{\rho}_{\mu} \nonumber,
\end{eqnarray}
where $\Omega_{\mu\nu}=\partial_{\mu}\omega_{\nu}-\partial_{\nu}\omega_{\mu} ,
\quad \mathbf{B}_{\mu\nu}=\partial_{\mu}\boldsymbol \rho_{\nu}-\partial_{\nu} \boldsymbol \rho_{\mu} - g_\rho (\boldsymbol{\rho}_\mu \times\boldsymbol{\rho}_\nu)$,  $m_\sigma$, $m_\omega$, and $m_\rho$ are the meson masses and $\kappa$, $\lambda$, $\zeta$,  $\xi$ and $\Lambda_\omega $ are
constant coupling parameters. We will consider models with constant couplings, NL, NL-$\omega\rho$, TM,  TM-$\omega\rho$, GM, H, FSU; the BSR models which include nonlinear meson terms, and DD models which have density dependent couplings, and for which the couplings related to nonlinear mesons terms are zero (in particular, the couplings $\kappa$, $\lambda$, $\xi$, $\zeta$ and $\Lambda_\omega$). For the DD models the isoscalar couplings of the mesons $i$ to the baryons take the form 
\begin{eqnarray}
g_{i}(n)=g_{i}(n_0)a_i\frac{1+b_i(x+d_i)^2}{1+c_i(x+d_i)^2} \, ,
\end{eqnarray}
and the isovector ones are given by
\begin{eqnarray}
g_{i}(n)=g_{i}(n_0)\exp{[-a_i(x-1)]} \, .
\end{eqnarray}
In these expressions, $n_0$ is the symmetric nuclear saturation density, and $x=n/n_0$.  For more information, we refer to references listed in the last column of Table \ref{tab:RMFmesons}.

For hyperonic matter, terms for naked hyperons and mesons mediating the hyperonic interaction are added to the Lagrangian density. Coupling constants for the YY and YN interactions are presented in Refs.~\cite{Fortin2017,Fortin2020,Providencia2019}. For the hyperonic sector, interactions are much simpler and include no cross terms between the hyperonic mesons \citep{Fortin2017}. For quark matter, terms for scalar, vector, and pseudovector quark couplings are added; we refer to Refs.~\cite{Pereira2016} and \cite{Ferreira:2020evu} for a discussion on the coupling constants in the quark Lagrangian. Once the Lagrangian density is established, the variation of the action (integral of the Lagrangian density) with regards to the wave functions and fields in play, yields a set of equations that must be solved numerically.

Another phenomenological approach to compute the EoS are Skyrme density functionals. NS matter is out of the reach of perturbative Quantum Chromo-Dynamics (QCD), as is oftentimes represented in QCD phase diagrams. However, mesons that mediate the strong interaction can be treated in an effective-field theory designed to replace the extension of QCD to low temperature neutron matter. The Skyrme force is a nonrelativistic approach to the interaction of nucleons; the Skyrme density functional derived from this interaction is  treated with the variational principle to design a Hamiltonian; baryons emerge as solutions to an approximation of Schr\"odinger equations (e.g., Hartree-Fock). In this framework, a series of (Skyrme) parameters ($x_{i}$, with $i\in[0,n]$; $t_{j}$, with $j \in [1,m]$; $W_0$ and $\alpha_k$ with $k \in [1,l]$) are used. The upside of this phenomenological approach is that some microscopic quantities (energy per baryon, effective nucleon mass, symmetry energy, incompressibility etc.) are established analytically from the above-mentioned parameters; for details, see the review \cite{Dutra2012}. Nuclei data tables and properties of homogeneous neutron matter are used to adjust the parameters. Skyrme density functionals can be separated in two classes: standard and generalized. To classify Skyrme models, we refer to the expression for the Skyrme force ruling the interaction between nucleons presented in Ref.~\cite{Goriely2010}: 

\begin{widetext}
\begin{align} \label{eq:skyrme}
    S_1(\bm{r}_{ij}) &= t_0 (1+x_0 P_{\rm s}) \delta(\bm{r}_{ij}) + \frac{t_1 (1+x_1 P_{\rm s})}{2 \hbar^2} \Big( \bm{p}_{ij}^2 \delta(\bm{r}_{ij}) + \delta(\bm{r}_{ij}) \bm{p}_{ij}^2 \Big) + \frac{t_2 (1+x_2 P_{\rm s}) }{\hbar^2}  \bm{p}_{ij} \cdot \delta(\bm{r}_{ij}) \bm{p}_{ij} \nonumber \\
    & + \frac{t_3 (1+x_3 P_{\rm s})}{6} \rho(\bm{r})^{\alpha_1} \delta(\bm{r}_{ij}) + \frac{ t_4 (1+x_4 P_{\rm s}) }{2 \hbar^2}\Big(  \bm{p}_{ij}^2 \rho(\bm{r})^{\alpha_2} \delta(\bm{r}_{ij}) + \delta(\bm{r}_{ij}) \rho(\bm{r})^{\alpha_2} \bm{p}_{ij}^2 \Big) + \frac{t_5 (1+x_5 P_{\rm s})}{\hbar^2}  \bm{p}_{ij} \cdot \rho(\bm{r})^{\alpha_3} \bm{p}_{ij} \nonumber \\
    &+ \frac{ {\rm i} W_0}{\hbar^2}(\bm{\sigma}_i + \bm{\sigma}_j) \cdot \bm{p}_{ij} \times  \delta(\bm{r}_{ij}) \bm{p}_{ij} \;. 
\end{align}
\end{widetext}
The quantity $\bm{r}_{ij}$ is defined as the difference between the spatial coordinates of nucleon $i$ and $j$, and $\bm{p}_{ij}$ designates the relative momentum (difference between the momentum operator of $i$ and of $j$). The spin-exchange operator between nucleons is denoted $P_s$, and $\rho(\bm{r}_{ij})$ is the local density or in other words the density at the barycenter $\bm{r} = 1/2(\bm{r}_i + \bm{r}_j)$. 

The terms of Eq.~\eqref{eq:skyrme} can be understood as follows: 
\begin{itemize}
    \item terms proportional to $t_0$ are the effect of the force at zero range (hence the $\delta$ function), 
    \item terms proportional to $t_1$ and $t_2$ are effects for an effective range, and express the momentum dependence of the interaction - consequently finite temperature effects; let us note that $t_4$ and $t_5$ represent the effective range in the generalized form of the Skyrme force, and introduce density dependence to the term, 
    \item terms proportional to $t_3$ account for a three-body interaction expressed as a density-dependent two body interaction, 
    \item terms proportional to $W_0$ account for the two-body spin interaction with spin-orbit coupling.
\end{itemize}
To the Skyrme force can be added a pair force and Wigner force as is the case for Brussels-Skyrme models presented in this paper.

Although it is computationally costly, microscopic EoSs can be established by solving the N-body Schr\"odinger equation. In this framework, the only requirement to construct the EoS is a solid understanding of the nucleon-nucleon interaction (calibrated to nuclear data). Different approaches exist, for example, the nonrelativistic Brueckner-Hartree-Fock approach or the relativistic Dirac-Brueckner-Hartree-Fock \citep{Taranto2013} approach. In practice, the N-body problem is reduced to a three-body problem, which is sometimes itself reduced to a density dependent two-body problem. Another promising microscopic approach which has been largely explored for neutron-star matter in recent years is Chiral Effective Theory, but it only reaches densities that are twice the saturation density; in practice, one can glue low density Chiral Effective Theory results to a higher density EoS, as is presented in e.g., Ref.~\cite{Sammarruca2022}.

\begin{table*}
    \begin{tabular}{|c|ccc|ccc|c|c|}
    \hline
    \multirow{2}{*}{Family} & \multicolumn{3}{l|}{Self-interaction} & \multicolumn{3}{l|}{Cross-interaction} & \multirow{2}{*}{Model} &  \multirow{2}{*}{Ref.} \\
    \cline{2-4}  \cline{5-7} 
    & $\sigma$ & $\omega$ & $\rho$ & $\sigma-\omega$ & $\sigma-\rho$ & $\omega-\rho$ &  &  \\
    \hline
    \hline
    \multirow{2}{*}{BSR} & \multirow{2}{*}{1} & \multirow{2}{*}{1} & \multirow{2}{*}{0} & \multirow{2}{*}{1} & \multirow{2}{*}{1} & \multirow{2}{*}{1} &  BSR2 & \multirow{2}{*}{\cite{Agrawal2010}} \\
    & & & & & & & BRS6 & \\
    \hline
    \multirow{3}{*}{DD} & \multirow{3}{*}{0} & \multirow{3}{*}{0} & \multirow{3}{*}{0} & \multirow{3}{*}{0} & \multirow{3}{*}{0} & \multirow{3}{*}{0} & DD2 & \cite{Typel2010} \\
    & & & & & & & DDME2 & \cite{Lalazissis2005} \\
    & & & & & & & DDH$\delta$ & \cite{GAITANOS2004}\\
    \hline
    \multirow{3}{*}{FSU} & \multirow{3}{*}{1} & \multirow{3}{*}{1} & \multirow{3}{*}{0} & \multirow{3}{*}{0} & \multirow{3}{*}{0} & \multirow{3}{*}{1} & FSU2 & \cite{Chen2014} \\
    & & & & & & & FSU2H & \cite{Negreiros_2018} \\
    & & & & & & & FSU2R &  \cite{Negreiros_2018}\\
    \hline
    \multirow{3}{*}{GM/H} & \multirow{3}{*}{1} &  \multirow{3}{*}{0} &  \multirow{3}{*}{0} &  \multirow{3}{*}{0} &  \multirow{3}{*}{0} &  \multirow{3}{*}{0}  & GM1 & first model of Table II in Ref.~\cite{Glendenning1991}\\
    &  &  &  &  & & & H3 &
    \multirow{2}{*}{\cite{Glendenning1991,Lackey06}}\\
    & & & & & & & H4 &  \\
    \hline
    NL & 1 & 0 & 0 & 0 & 0 & 0  & NL3 & \cite{Lalazissis1997} \\
    \hline
    NL-$\omega\rho$ & 1 & 0 & 0 & 0 & 0 & 1 & NL3$\omega\rho$ & \cite{Horowitz2001} \\
    \hline
    \multirow{2}{*}{TM} & \multirow{2}{*}{1} & \multirow{2}{*}{1} & \multirow{2}{*}{0} & \multirow{2}{*}{0} & \multirow{2}{*}{0} & \multirow{2}{*}{0}  & TM1 & \multirow{2}{*}{\cite{SUGAHARA1994}} \\
    & & & & & & & TM2 &  \\
    \hline
    \multirow{2}{*}{TM-$\omega\rho$} & \multirow{2}{*}{1} & \multirow{2}{*}{1} & \multirow{2}{*}{0} & \multirow{2}{*}{0} & \multirow{2}{*}{0} & \multirow{2}{*}{1}  & TM1$\omega\rho$ & \multirow{2}{*}{\cite{Providencia2013} }\\
    & & & & & & & TM2$\omega\rho$ & \\
    \hline
  \end{tabular}
  \caption{Classification of relativistic mean-field models, by the interacting mesons that appear in the Lagrangian density of the theory.} \label{tab:RMFmesons}
\end{table*}


In this paper, we use the following cold matter EoSs: 
\begin{itemize}
    \item Nucleonic relativistic mean-field (x15): 
    \begin{itemize}
        \item BSR2, BSR6,
        \item DD2, DDME2, DDH$\delta$: in this family of EoSs, no self or cross interactions are taken into account, but the meson coupling constant in the covariant derivative and effective mass of the naked baryon are density dependent,
        \item FSU2, FSU2H, FSU2R,
        \item GM1,
        \item NL3, NL3$\omega \rho$,
        \item TM1, TM1$\omega \rho$,
        \item TM2, TM2$\omega \rho$;
    \end{itemize}
    for references, see Table.~\ref{tab:RMFmesons}.
    \item Hyperonic relativistic mean-field (x7): for references, see \cite{Fortin2017,Fortin2020}.
    \begin{itemize}
        \item DD2 and DDME2,
        \item FSU2H,
        \item H3, H4: based on the same nuclear model as the purely nucleonic GM1. The parameters for the nucleonic sector of this seminal model were chosen to reproduce simple nuclear constraints on the incompressibility ($K=300$ MeV) and effective mass at saturation ($m^\ast/m=0.7$). Much more up-to-date and refined models are now available and used in this paper. We however employ GM1, H3, and H4 in this work as a comparison with \cite{Read09}. Note that while GM1 and H4 have maximum masses larger than $2\,M_\odot$, H3 does not: $M_{\rm max}\simeq 1.79$\;M$_{\odot}$, see Table.~\ref{tab:rmfmacro},
        \item NL3, NL3$\omega \rho$.
    \end{itemize}
        
    \item  Hybrid relativistic mean-field (x5): for references, see Refs.~\cite{Pereira2016} and \cite{Ferreira:2020evu}.
    \begin{itemize}
        \item DD2-B15-40-20
        \item NL3$\omega\rho$B20-50-0,  NL3$\omega \rho$-B28-75-0
        \item NL3$\omega \rho$-B0-50-0
        \item NL3$\omega \rho$-B0-50-50
    \end{itemize}
    The terms for mesons self and cross interaction in the Lagrangian density are used to categorize the different family of models in Table~\ref{tab:RMFmesons}. Within a family of EoS, the difference between models lies in the value of the coupling constants. 
    \item Nucleonic Skyrme (x24):
    \begin{itemize}
        \item BSk20, BSk21, BSk22, BSk23, BSk24, BSk25, BSk26, 
        \item KDE0v1,
        \item Rs,
        \item Ska, Skb, SkOp, SkMP, Sk255, Sk272,
        \item SkI2, SkI3, SkI4, SkI5, SkI6, 
        \item SLy2, SLy9, SLy230a,
        \item DH: based on SLy4 parametrisation but contrary to the other SLy EoSs presented in this paper, it has been calculated consistently for the core and the crust;
    \end{itemize} 
    Classification of Skyrme EoSs and references are presented in Table~\ref{tab:SkyrmeEOS} using the Skyrme parameters presented in Eq.~\eqref{eq:skyrme}.
    \item \textit{ab-initio} (x1): BCPM as presented in Ref.~\cite{Sharma2015}. It is based on the microscopic Brueckner-Hartree-Fock theory of nucleon interaction with the addition of a density dependent two-body force.
\end{itemize}
All above-mentioned EoSs with the exception of H3 meet the maximum mass criterion. 

\begin{table*}
\begin{tabular}{|c|c|c|c|}
\hline
Family & Parameters & Model & Ref. \\
\hline \hline
\multirow{2}{*}{BSk} & \multirow{2}{*}{$t_2=0, t_2x_2 \ne 0$} & BSk-20/21 & \cite{Goriely2010} \\
 & & BSk-22/23/24/25 & \cite{Goriely2013} \\
\hline
\multirow{3}{*}{SLy} & \multirow{3}{*}{$x_4=x_5=t_4=t_5 = 0$} & SLy-2/9 & \cite{Chabanat1995} \\
 & & DH & \cite{Douchin01} \\
 & & SLy230a & \cite{Chabanat1998} \\
\hline
KDE & $t_4 = t_5 = 0$ & KDE0v1 & \cite{Agrawal2005} \\
\hline
Rs & $x_1 = x_2 = t_5 = 0$ & & \cite{Friedrich1986} \\
\hline
Sk & $x_1 = x_2 = x_4 = x_5 = t_4 = t_5 = 0$ & Sk-a/b &\cite{Kohler1976} \\
\hline
\multirow{5}{*}{Sk}& \multirow{5}{*}{$x_4=x_5=t_4=t_5=0$} & Sk-255/272 & \cite{Agrawal2003} \\
& & SkMP & \cite{Bennour1989} \\ 
& & SkOp & \cite{Reinhard1999} \\ 
& & SkI-1/2/3/4/5 & \cite{Reinhard} \\
& & SkI6 & \cite{Nazarewicz1996} \\
\hline
\end{tabular}
\caption{Classification of Skyrme models from parameters presented in Eq.~\eqref{eq:skyrme}. In column one is the Skyrme family, in column two, are presented the parameters of the Skyrme force that are included in the model, in column three is the name of the EoS and in column four is the reference to the model.}
\label{tab:SkyrmeEOS}
\end{table*}


\section{Polytropic fits}\label{PP}

In the following, convention as presented in Ref.~\cite{Read09} (later referred to as PPFRead with PPF standing for Piecewise Polytropic Fits) is used: the rest-mass density $\rho$ is directly connected to the baryonic density $n$ via the baryon mass $m_B =939$\;MeV/c$^2$ ($\rho=n\,m_B$). The energy density $\epsilon$ is calculated by using the first law of thermodynamics in the zero-temperature limit:
\begin{equation} \label{eq:thermo}
    d\bigg(\frac{\epsilon}{\rho} \bigg) = - P d\bigg(\frac{1}{\rho}\bigg) \;. 
\end{equation}

\subsection{Piecewise polytropes}
A polytrope takes the form 
\begin{equation} \label{eq:poly}
    P = \kappa \rho^{\Gamma}, 
\end{equation}
with $\kappa$ being the polytropic constant, and $\Gamma$ being the adiabatic index. This type of crude approximation for the EoS is simple but oftentimes used: for example, the outer crust of NSs is well approximated by the pressure of ultrarelativistic electrons ($\Gamma = 4/3$). The whole NS cannot be accurately approximated with only one polytrope; however, it is  possible to divide the EoS in $N$ parts, each of which would pertain to a polytrope with fixed $\kappa$ and $\Gamma$, which is what we call piecewise polytropes. 

A practical fit contains a restricted number of polytropes in order for the parametrisation to be convenient. PPFRead are based on $N=7$ polytropes with four in the crust and three in the core and we shall use it as well. One could suppose that a total number of $3\times N -1$ parameters is needed to fit the EoS: $N$ $\Gamma$s, $N$ $\kappa$s and $N-1$ transition densities $\rho_{t[i\to i+1]}$ that defines areas of the EOS by which polytrope $i$ is fitted (with $i \in [1,N]$). However, pressure continuity reduces the number of parameters to $2 \times N$ with $N$ adiabatic indices, $N-1$ transition densities and only one $\kappa$ : at $\rho_{t[i\to i+1]}$, $\kappa_{i+1}$ is calculated from $\Gamma_{i}$, $\kappa_i$, and $\Gamma_{i+1}$. 

Integrating Eq.~\eqref{eq:thermo} with the help of Eq.~\eqref{eq:poly} gives an expression for the energy density that depends on the adiabatic index and polytropic constant : 
\begin{equation}\label{eq:Energydensity}
    \epsilon(\rho) = (1+a_i)\rho + \frac{\kappa_i}{\Gamma_i -1}\rho^{\Gamma_i}, 
\end{equation}
with $a_i$ being the integration constant determined at the transition between polytropes. Expression for this constant can be found in Ref.~\cite{Read09}, and its initial value is a physical requirement: the rest mass density zero limit implies $\epsilon = \rho$. 

The major downside of piecewise polytropic fits is the nonderivability of quantities; for example, the sound velocity is not continuously defined. 

\subsection{Highlight on the importance of unified equations of state}

It is common to find nonunified EoSs in the literature: the core and the crust EoS are not computed using the same nuclear model. However, nonunified constructions may induce non-negligible errors in the modeling of macroscopic parameters. Those constructions exist because computing the crust inhomogeneities is more tedious than computing the homogeneous core. A widespread practice within the astronuclear physics community, is to compute a core EoS, and glue an already established crust EoS. If this core-crust matching is performed with no care for thermodynamic consistency, it can lead to nonphysical jumps in the pressure. Only part of the outer crust is constrained by nuclei data, such that the high density outer crust and the inner crust are subject to model dependent differences. Therefore, core-crust matching of two EoSs with very different microscopic parameters such as the symmetry energy $J_m$ and its slope $L_m$ at the matching density, results in errors as large as 20\% in the modeling of macroscopic parameters, for details see Ref.~\cite{Suleiman2021}. 

The relativistic hydrodynamics equations are nonlinear and the NS interior is opaque to observations. Therefore, one cannot separate the contribution of different parts of the EoS in the modeling of the total macroscopic parameters. Let us note that some quantities are more sensitive to certain parts of the star interior: for example the role of the crust treatment is more important for the radius of the star than for the mass, the total moment of inertia and the tidal deformability. If one wants to explore high density matter with NS observations, the artificial errors introduced by nonunified models can mislead into the acceptance or exclusion of the investigated nuclear model. 

Examples for the use of nonunified EoSs are found in more simulations than can be listed: analytical representations used for GW data, finite temperature simulations, modelization of NS parameters in modified gravity, magneto-hydrodynamics, "universal" relations, etc.

All EoSs presented in Sec.~\ref{sec:allEOS} are unified models: 
\begin{itemize}
    \item The set of EoSs based on RMF models are taken from Refs.~\cite{Fortin16,Fortin2020,Providencia2019}; those that include hyperons are constructed consistently with the available experimental measurements of the properties of hypernuclei Refs.~\cite{Fortin2017,Fortin2020}. The EoS for the inner crust is calculated within the Thomas-Fermi approximation consistently with the EoS of the core \cite{Grill2014,Providencia2019}. The EoS of the outer crust has not been obtained consistently but taken from Ref.~\cite{Ruster}. Several other outer crust EoSs are available (\textit{e.g.}, Refs.~\cite{BP,HP}) but they all are strongly constrained by nuclear physics data and therefore very similar. In addition, as mentioned in Ref.~\cite{Fortin16}, we have checked that the  use of an EOS for the outer crust not fully consistent with the rest of the star does not significantly affect the star properties, the radius in particular, for masses above 1.0\;M$_\odot$, which is the mass range of all NS currently observed.
    \item The set of EoSs based on Skyrme models are taken from Ref.~\cite{Fortin16}. The construction of the crust does not include shell effects and curvature terms which results in a mass shift with respect to experimentally measured masses, see Ref.~\cite{Gulminelli2015}. Thus the EoS of the external part of the outer crust differs from the one we would get by employing experimental data but it is small enough to impact the relation between the mass and the radius by less than 1\%. 
    \item For the \textit{ab initio} EoS, the outer crust is based on the formalism of Ref.~\cite{BP}; a density functional designed from Brueckner-Hartree-Fock computations is used in the deformed Hartree-Fock-Bogoliubov formalism for nuclei not included in the data table \cite{ame2012}. For the inner crust, the energy density functional derived from Brueckner-Hartree-Fock calculations is used in the Thomas-Fermi approximation.
\end{itemize}

A few groups care for the unification of EoSs in a different manner, for example, Ref.~\cite{Chatterjee_2017} proposes a meta model paired to the extended Thomas Fermi approximation for the low density part of the EoS. References \cite{Fereira2020,Ferreira2020b,Ferreira2021} 
use a set of causal model-independent EoS obtained from a Taylor expansion around the saturation density; at low density, a crust EoS is matched in a thermodynamically consistent way.

\begin{figure}
\resizebox{\hsize}{!}{\includegraphics[scale=0.3]{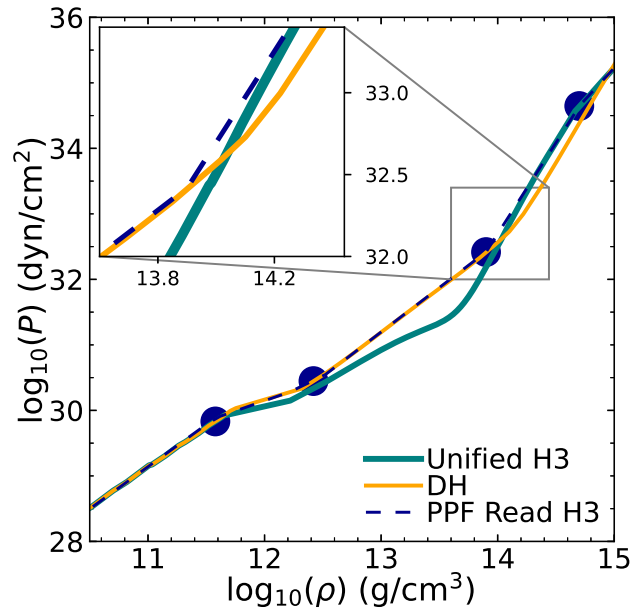}}
\caption{Equation of state (relation between the pressure $P$ and the density $\rho$) for unified EoS H3 compared with Read's piecewise polytropic fit of H3 (PPFRead). Transition between polytropes of PPFRead are presented with blue points. }
\label{fig:CompReadeos}
\end{figure}

PPFRead are reduced to a core fit, to which the fit of EoS DH's crust is attached at a matching transition pressure which is not derived from either crust or core nuclear model. Their construction is the following: 
\begin{itemize}
    \item the high density part of the EoS is fitted by three polytropes;
    \item the low density part of the EoS is fitted by four polytropes based on EoS DH;
    \item the point at which the high density and the low-density polytropes are matched is adapted to ensure a minimal fit error on the whole nonunified construction; 
    \item the last polytrope of DH is prolonged or shortened to ensure that it crosses the first polytrope of the high-density part.
\end{itemize}
Three models of our set of EoSs overlap with Read's fitted ones: DH, H3 and H4. We investigate the following constructions:
\begin{itemize}
    \item the unified EoS H3 and H4,
    \item the unified EoS DH, 
    \item PPFRead for H3, H4 and DH,
\end{itemize}
and present the relation between pressure and baryonic density in Fig.~\ref{fig:CompReadeos}. Results are not presented for EoS H4 in this figure because our focus is on the low-density part of the star, and H3 and H4 diverge only in the core (the difference between H3 and H4 lies in the hyperonic meson couplings). The lowest-density parts of the EoSs, that is to say, $\rho < 10^{11.6}$\;g/cm$^{3}$, overlap for all constructions: this corresponds mostly to the outer crust which is calibrated to experimental data; therefore, it is similar for all nuclear models. Over $10^{11.6}$\;g/cm$^{3}$, DH and unified H3 are different. The zoom in the figure highlights the matching area: for H3, the matching density is ${\rho^{\rm H3}_{m} = 7.9477 \times 10^{13}}$\;g/cm$^3$; for H4, ${\rho^{\rm H4}_{m} = 8.8774 \times 10^{13}}$\;g/cm$^3$. In between the last point of convergence for all constructions, and the matching of DH and H3 PPFRead, the curves are different which attests to the limit of laboratory measurement calibrations of the low density part of the EoS. Although precautions are taken to avoid jumps in the pressure, the differences between the sole DH crust and the various core EoSs in PPFRead are not negligible.

\begin{figure*}
\resizebox{\hsize}{!}{\includegraphics[scale=0.3]{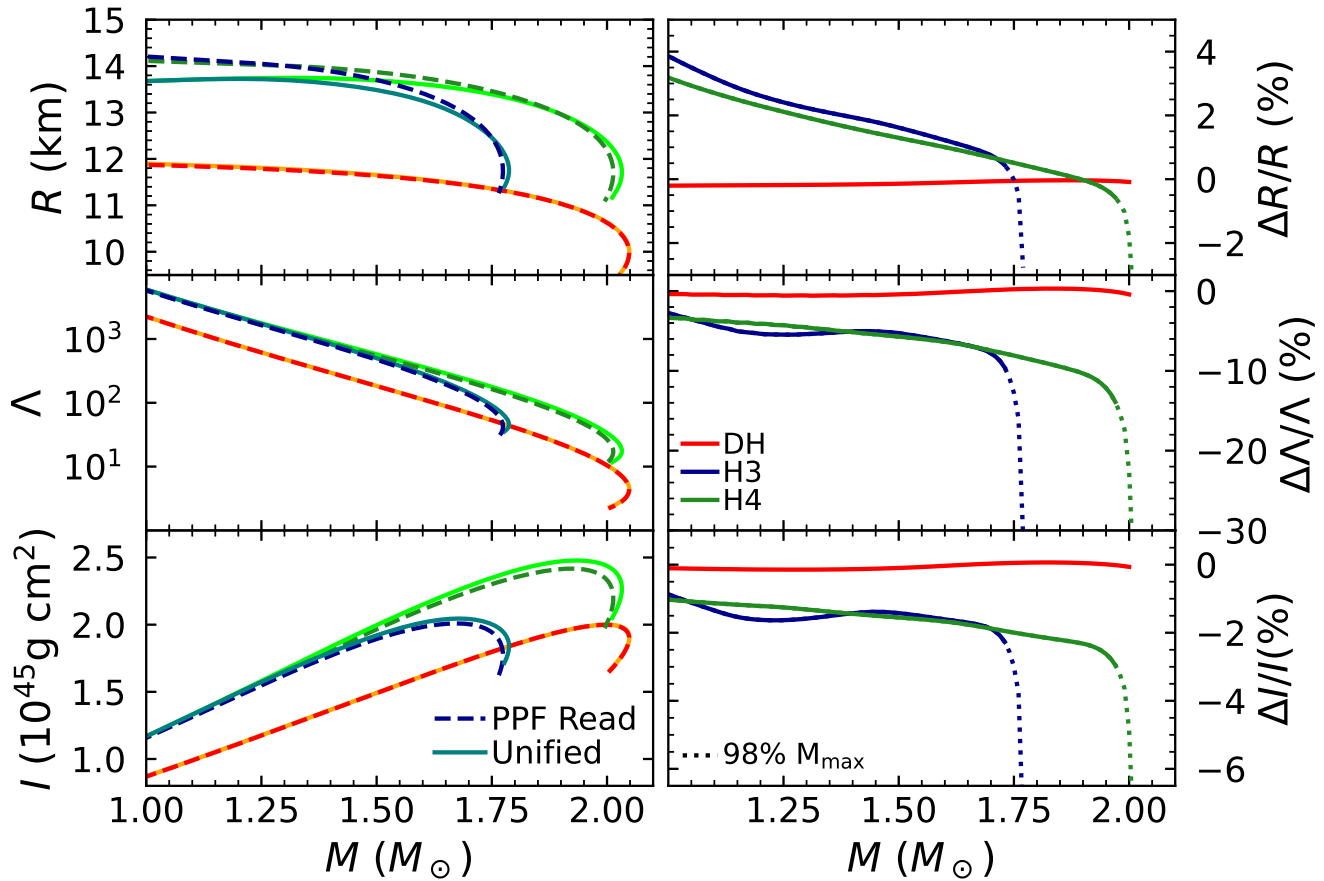}}
\caption{Radius $R$, tidal deformability $\Lambda$, and moment of inertia $I$ as a function of the mass $M$ for unified tables H3, H4, and DH, as well as Read's piecewise polytropic fits (PPFRead). In the right panel, the relative difference at given mass $M$ between unified tables and PPFRead for each macroscopic parameters in the three cases of EoS is presented. For EoS H3 and H4, the relative uncertainty is shown up to 98\% of the maximum mass in plain lines, and the last two percent in dotted lines, see Sec.~\ref{sec:ppfperform} for details.}
\label{fig:CompRead}
\end{figure*}

From the above mentioned constructions, macroscopic parameters of a cold and non-rotating NS are computed using the relativistic equations of hydrostatics, the tidal deformability and the moment of inertia equations. Results for the radius $R$, dimensionless tidal deformability $\Lambda$, and moment of inertia $I$ as a function of the mass $M$ are presented in Fig.~\ref{fig:CompRead} for unified H3, H4, and DH as well as their respective PPFRead. The relative difference between the fits and unified tables for the variable $X$ at given mass $M$ is given by 
\begin{equation} \label{eq:delta}
    \Delta \equiv \Delta X/X = \big(X_{\rm fit}(M) - X_{\rm uni}(M) \big)/X_{\rm uni}(M) \;,
\end{equation}
and is presented in the right panels of Fig.~\ref{fig:CompRead}. DH PPFRead is constructed from DH low density polytropes and DH high density polytropes, therefore is unified: the red line in the right panel of Fig.~\ref{fig:CompRead} shows that it coincides with the tabulated unified EoS DH. We conclude that the fit method is powerful when applied to unified EoSs. In the case of PPFRead H3 and H4, the low density polytropes of DH are matched to the high density polytropes of H3 and H4 at $\rho^{\rm H3}_{m}$ and $\rho^{\rm H4}_{m}$. Therefore, the whole fit is not unified and differs from our unified tables as is shown for green and blue lines in Fig.~\ref{fig:CompRead}. This indicates that using nonunified EoSs for piecewise polytropic fits induces an artificial error on the macroscopic parameters. 

From Fig.~\ref{fig:CompRead}, the uncertainty related to the tidal deformability leads to two interesting points. On the one hand, despite the largest relative error being that of the tidal deformability, in the left panels of Fig.~\ref{fig:CompRead} the accuracy looks very similar for all three quantities. This is due to the fact that $\Lambda$ is plotted on a logarithmic scale, because it spans over two orders of magnitude in the interesting range of masses. As a consequence, the relative change of $\Lambda$ by $\approxeq 15\%$ corresponds to $\approxeq 2\%$ in mass and such error bars would be of a similar size in Fig.~\ref{fig:CompRead} (left panel). On the other hand, one can notice that the sign of the relative difference in radius and that in tidal deformability are not the same. However, $\Lambda$ is calculated according to the relation ${\Lambda=\frac{2}{3} k_2 C^{-5}}$ \citep{De2018, Malik18}, with $k_2$ being the tidal Love number solved simultaneously with the Tolman-Oppenheimer-Volkoff equations. The dimensionless deformability being proportional to $k_2 R^5$, an increase in radius should correspond to an increase in the tidal deformability. The sign difference in the relative uncertainty of the two quantities in Fig.~\ref{fig:CompRead} can be explained by the large error of PPFRead with regards to the unified EoS on the quantity $k_2$. It is particularly large for low mass stars, such that it dominates the $R^5$ factor. For higher mass stars, the uncertainty on $k_2$ is smaller and the scale of $R^5$ dominates. This large $k_2$ error can be understood as the strong effect of the crust matching on this quantity. Indeed, the fit from H3 unified tables in which the crust is treated correctly with the core -as is presented in details in the next section- gives a relative error on $k_2$ which is at most 3\% (for a 1.5\;M$_{\odot}$) whereas PPFRead give a relative error on $k_2$ of at most 50\% (for a 1.0\;M$_{\odot}$).

In the next section, we provide piecewise polytropic fits parameters based on unified tables of EoSs presented in Sec.~\ref{sec:allEOS}.

\subsection{Method for the fit}

To revise PPFRead with unified tables of EoSs, we use an adaptive nonlinear least squares method to calculate the $N_{\rm poly}-1$ transition densities $\rho_t$, with the number of polytropes $N_{\rm poly}=7$. They are adapted such that the fit error is minimized for the entire EoS (core and crust) of the $P(\rho)$ fit.

For each unified table, we create a distribution for the number of density points $\rho$ different from the original tabulated EoS. Our distribution of points allows us to give more importance to the accuracy of the fit in some parts of the star than others. We use a total of 1500 points, allocating 1/5 of the points in the crust and 4/5 of the points in the core, because it resulted in better fits of the full EoS. In each region, we distribute these points uniformly on a logarithmic scale. 
The fragment of the unified tables with the largest densities is not used because the density goes beyond the central density at maximum mass.
We eliminate this highest density part by calculating the maximum mass of the star from the unified table, and only interpolate the EoS up to $n_{\rm max}$, thus increasing the fit accuracy for astrophysical quantities. 

The core-crust transition density has a particular influence on the success of our fit method. This area is particularly sensitive to changes in the polytropic parameters and in turn, the points could be miss-allocated. Therefore, we chose to test values of the core-crust transition densities from $n=0.06$ to $n=0.14$\;fm$^{-3}$ for each EoS fit. 

The relation $P(\rho)$ is interpolated by using a first-order spline method to establish the pressure points from our distribution of density points. Each polytrope is fit by using a nonlinear least squares method from Eq.~\eqref{eq:poly}; $\Gamma$ and $\kappa$ are determined for each polytrope. Then, transition densities are recalculated from the polytropic parameters just fitted and the whole process with these new transition densities starts over until the set of $\rho_t$ stagnates. Finally, the energy density $\epsilon$ is calculated from Eq.~\eqref{eq:Energydensity}. 

Some alternatives to this fit method have been explored. With regards to the number of points for the interpolation, we have tested values between 200 and 10 000 points and observed a plateau of accuracy for $\approxeq$1500 points. We have also tested an inverse fit method, starting the fit from high density to low density which renders a similar accuracy.

The method described above is implemented to calculate the fit parameters which are presented in Tables \ref{tab:RMFparam}, and \ref{tab:skyrmeParam} in the Appendix. We provide a routine in \textit{Python} and tables of fit parameters in ASCII format to compute the tabulated fitted EoSs presented in this paper, see Ancillary files.

\section{Role of unified EoSs in the accuracy of macroscopic parameters} \label{macro}

\subsection{Piecewise polytropic fit's performance} \label{sec:ppfperform}

The method we use to establish piecewise polytropic fits of unified EoSs is intended to provide accurate modeling of macroscopic parameters for a nonrotating NS. Therefore, quantities $M$, $R$, $I$, and $\Lambda$ are calculated from our fitted EoSs and compared with that of unified tables. Results for macroscopic parameters of prime interest are presented in Table~\ref{tab:rmfmacro}, and \ref{tab:Skyrmemacro}: the maximum mass $M_{\rm max}$, density at the maximum mass $n_{\rm max}$, the radius at 1.0\;M$_{\odot}$ and 1.4\;M$_{\odot}$ respectively denoted $R_{1.0}$ and $R_{1.4}$, as well as the radius at maximum mass $R_{\rm max}$, the moment of inertia at $1.338$\;M$_{\odot}$ denoted $I_{1.338}$ and at the maximum mass $I_{\rm max}$, and the tidal deformability at 1.4\;M$_{\odot}$ denoted $\Lambda_{1.4}$ and at maximum mass $\Lambda_{\rm max}$. We provide the relative errors $\Delta$ on these quantities except those defined at the maximum mass. For the latter we include instead the relative error $\delta$ defined as the relative difference between the quantities calculated at the maximum mass for the unified table, and at the maximum mass for our fit. Indeed since the maximum mass of the unified EoS and our fit are not equal, $\Delta$ at the maximum mass and $\delta$ are different. The largest errors are presented in red in the tables.

\begin{figure}
    \centering
    \includegraphics{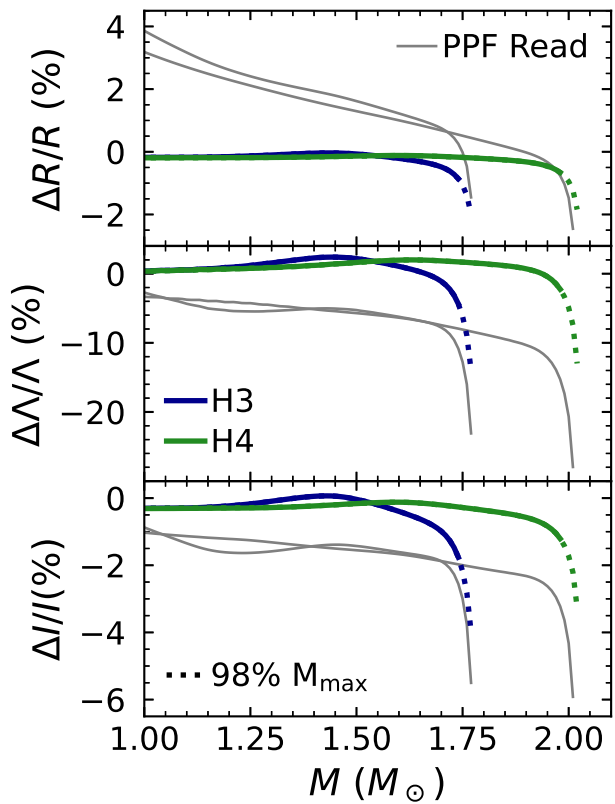}
    \caption{Relative difference at given mass $M$ in pourcent between unified tables and our fits for the radius $\Delta R/R$, the tidal deformability $\Delta \Lambda /  \Lambda$ and the moment of inertia $\Delta I/I$ as a function of the mass $M$ for EoS H3 and EoS H4. The relative difference between PPF Read and unified EoSs H3 and H4 is presented in gray. For our fits, the relative difference is presented up to 98\% of the maximum mass in plain lines, and the last two percent in dotted lines, see text for details. }
    \label{fig:Ourfit}
\end{figure}

With regards to nucleonic relativistic mean-field models, errors associated with the fit on $M_{\rm max}$, $n_{\rm max}$, and quantities related to the radius are systematically under under 1\%. For astrophysical quantities related to the moment of inertia, it stays under 1.5\%, and for the tidal deformability under 4\%. For hyperonic relativistic mean-field models EoS consistent with the maximum mass criterion (all but H3), the errors associated to the fit on $M_{\rm max}$, $n_{\rm max}$, and quantities related to the radius stay under 1.5\%. Quantities related to the moment of inertia stay under 2.5\%, and for the tidal deformability under 7\%. For hybrid relativistic mean-field models, the errors associated with the fit stays under 2\% for all quantities.

For Skyrme models, the error associated with the fit for $M_{\rm max}$, $n_{\rm max}$, quantities related to the radius and the moment of inertia are under 1\%. Once again, the tidal deformability does not fair as well, with an error up to 5\%.  

The maximum mass is the most accurately reproduced quantity, with an error under 0.5\%. The tidal deformability is systematically the quantity with largest errors associated with the fit. Generally, nucleonic models are more accurately reproduced by our fits than hyperonic or hybrid ones. This is understandable because the number of polytropes chosen in the core is fixed to three, and the presence of hyperons or a phase transition to deconfined quarks in the core produces respectively an additional softening and drop of the adiabatic index as a function of the baryonic density (see Fig.~\ref{fig:gammaAll} in the Appendix).

In Fig.~\ref{fig:Ourfit}, the relative difference at given mass $M$ between our fit and the unified tables for the radius, the tidal deformability and the moment of inertia are presented as a function of the mass for EoSs H3 and H4. The same quantities are plotted in gray for PPFRead to compare. The impact of using unified fits on the relative difference for the radius is particularly important for low mass stars, because the crust is relatively more significant for such objects.
In overview, the relative difference for our fit is significantly smaller than for PPFRead for all macroscopic parameters considered.

The large increase of the relative error close to the maximum mass (Fig.~\ref{fig:CompRead} right panel and Fig.~\ref{fig:Ourfit}) is a consequence of choosing the mass as an independent variable. Close to the maximum mass point, the error increases significantly even for a small difference between the values of the maximum mass for the original and fitted EoS. This effect is absent for the dependence of the error as a function of central density or pressure. The relative inaccuracy of our fits for stellar configurations with the same central pressure is smaller in particular in the region close to maximum mass and for $R$, $\Lambda$ and $I$ are 0.1\%, 1.8\%, 1\% respectively, compared with 1\%, 8.7\%, and 2\% for a fixed mass $M=2$\;M$_{\odot}$ for EoS H4.

In overall, our fits perform well beyond the expected current precision from NICER for the radius and for the mass. 

\begin{table*}
\caption{Maximum relative difference in \% between the compactness calculated via fits of universal relations established by Godzieba et al. and Yagi \& Yunes, and various constructions of EoS discussed in this paper for NL3, BSR6 and DD2. In parentheses, we present the value of the mass at which the maximum error is calculated.}
\label{tab:radice}
\centering
\begin{tabular}{c|cccc|cccc}
\hline 
\hline 
EOS & uni. & $0.15n_0$ & $0.1$fm$^{-3}$ & $1.2n_0$ & uni. & $0.15n_0$ &  $0.1$fm$^{-3}$ & $1.2n_0$ \\ 
\hline 
$C-\Lambda$ & \multicolumn{4}{c|}{Godzieba et al.} & \multicolumn{4}{c}{Yagi \& Yunes} \\ 
\hline
NL3 & 2.37 (1.54) &  7.14 (1.01) &  2.88 (1.40) &  2.90 (2.77) &  2.94 (1.01) &  4.66 (1.28) &  1.78 (1.00) &  2.78 (2.47) \\
BSR6 & 3.77 (1.21) &  6.85 (1.00) &  3.98 (1.01) &  2.28 (1.01) &  2.12 (1.33) &  4.53 (1.20) &  2.20 (1.31) &  1.20 (2.22) \\
DD2 & 2.78 (1.00) &  4.84 (1.01) &   3.53 (1.01) &   2.70 (1.01) &   1.11 (2.22) &  2.66 (1.17) &  1.52 (1.23) &  1.34 (2.21) \\
\hline
H3 & 5.80 (1.79) &  7.83 (1.79) &   6.60 (1.79) &   5.35 (1.78) &   4.20 (1.79) &  6.33 (1.79) &  5.02 (1.79) &  3.76 (1.78) \\
H4 & 3.64 (2.03) &  7.72 (1.00) &   4.28 (2.03) &   3.17 (2.03) &   3.09 (1.00) &  5.44 (1.20) &  3.54 (2.03) &  2.49 (2.03) \\

\hline
\hline
\end{tabular} 
\end{table*}

\subsection{Universal relations}
The story of "universal" relations started when \cite{Yagi2013} established the famous I-Love-Q (moment of inertia, tidal deformability and quadruple moment) relations; the quadruple moment $Q$ expresses the deformation of the external gravitational field of a star by its rapid rotation (hence nonsphericity). It appeared that relations between some macroscopic parameters was only minimally EoS dependent. These relations have powerful predictive power, for the measurement of one parameter would permit the extraction of the other: when the first I-Love-Q relations were established, there was hope to extract the tidal deformability from the moment of inertia, when actually it is the other way around thanks to the detection of GWs. Many other relations appeared throughout the years and were used in various NS simulations. The physical meaning of this universality has been attributed to two main reasons: the first being the low density dependence of some of those relations and the second being the extension of the no-hair theorem\footnote{The no-hair theorem refers to a theorem in general relativity, stating that all properties of a black hole depend solely on its mass, angular momentum and electric charge. This theorem does not necessarily hold in modified gravity, in which black hole can be hairy e.g., see Ref.~\cite{Gervalle_2020}.}. The former refers to the calibration to laboratory experiments of the low-density part of the crust, but we have discussed that this argument is viable for part of the outer crust only. The latter argument is discussed in length in Refs.~\cite{Yagi2014, Sham2015}: the authors explore the approximate baldness of NSs, and the analytical meaning of an extrapolated no-hair theorem in general relativity. It is suggested that this universality may rise from an emergent symmetry, acquired when the EoS parameters of a star are tuned out from main sequence, to relativistic stars, to black holes for which universality perfectly holds as per the no-hair theorem. 

Universal relations have been established by fitting the modelled macroscopic parameters calculated from existing EoSs which are nonunified. It has been shown in Ref.~\cite{Suleiman2021} that the precision presented for some of those relations no longer held when compared with the modeled relation of parameters calculated with unified EoSs. Therefore, although the quasi-universality of those relations is not put into question, the precision of the fits can be tainted by the use of nonunified EoSs. 

Recently, Godzieba et al. \cite{Godzieba2021} proposed a revision of universal relations for multipole Love numbers based on the fit of polytropic type constructions. Under approximately the saturation density, the EoS is that of DH approximated by one polytrope, whereas, in the core, three polytropes are used with parameters adjusted to a Monte-Carlo-Markov-Chain: around two million EoSs are created, following basic rules of causality, maximum mass constraint around 2\;M$_{\odot}$, and GW170817 measurement of the tidal deformability; this construction of the EoS is not unified, however, authors use a method that adapts the matching density between [0.15-1.2]\;n$_0$ at the junction of DH and their polytropic cores, similarly to PPFRead. Universal relations are established for dimensionless electric tidal deformability $\Lambda$ of order two, three and four with the compactness, but we focus on the relation $C-\Lambda_2$ which relates the compactness and the tidal deformability as measurable by GW detectors as of today. The fit is a logarithmic expansion as presented by \cite{Yagi17} (Yagi \& Yunes) and by Ref.~\cite{Maselli13} (Maselli et al.): 
\begin{equation}
    C_{\rm fit}=\sum_{k=0}^{N}a_k(\ln\Lambda)^k \;.
\end{equation}

The fit of Godzieba et al. yields ${a_0 = 0.3388}$, ${a_1= -2.3 \times 10^{-2}}$, ${a_2 = -4.651 \times 10^{-4}}$, ${a_3 = -2.636 \times 10^{-4}}$, ${a_4 = 5.424 \times 10^{-5}}$, ${a_5 = -3.188 \times 10^{-6}}$ and ${a_6 = 6.181 \times 10^{-8}}$. We calculate the relative difference between the compactness calculated with various EoS constructions and the compactness calculated using Godzieba et al., Yagi \& Yunes, and Maselli et al. fits: 
\begin{equation}
    \Delta C_{\rm fit} = \frac{|C-C_{\rm fit}|}{C_{\rm fit}}\;.
\end{equation}
The following EoS constructions are used:
\begin{itemize}
    \item unified EoS NL3 (stiff) and DD2 (soft), and BSR6 (in between),
    \item DH crust matched to NL3, DD2, BSR6, H3 and H4 core at 0.15$n_0$, 
    \item DH crust matched to NL3, DD2, BSR6, H3 and H4 core at 0.1 fm$^{-3}$, density at which \cite{Suleiman2021} have shown that the uncertainty due to the use of nonunified model is minimized for models NL3, DD2 and BSR6, 
    \item DH crust matched to NL3, DD2, BSR6, H3 and H4 core at 1.2$n_0$.
\end{itemize}
The relation $C-\Lambda_2$ established from unified EoSs, and the universal relation fit of Godzieba et al., Yagi \& Yunes, and Maselli et al. are presented in the appendix in Fig.~\ref{fig:Radice} for NL3 and DD2, and in Fig.~\ref{fig:Radice2} for H3 and H4. For both EoS NL3 and DD2, the quasi-universality of the relations is evident from the bottom plots of Fig.~\ref{fig:Radice}; all three universal relation fits overlap the relation $C(\Lambda)$ for unified EoSs. In the case of H3 and H4 whose results are presented in Fig.~\ref{fig:Radice2}, the deviation from the universal relations is visible, particularly for high mass NSs. After calculating the accumulated difference on all three fits of universal relations and unified EoSs, we found that the Yagi \& Yunes one fairs better than Maselli et al., itself faring slightly better than Godzieba et al. for EoSs NL3, DD2, H3 and H4. 

Additional results are presented in Table~\ref{tab:radice}: the maximum relative difference between the fit for universal relations, and the EoS constructions is displayed with the mass at which it is calculated. We choose to show results only for Yagi \& Yunes fit and Godzieba et al. fits, because Maselli et al. was shown to step outside of their reported error when compared with unified tables, see Ref.~\cite{Suleiman2021}. We find that the Yagi \& Yunes fit is coherent with its 6.5\% reported error, for all constructions of EoSs. In Ref.~\cite{Godzieba2021}, the authors emphasize that their fit is an improvement on Yagi \& Yunes one based on their collection of two million polytropic EoS constructions, however, when compared with tables of unified EoSs, the fit of Godzieba et al., which is based on nonunified constructions, generally fairs worse than the fit of Yagi \& Yunes. 

In overview, Yagi \& Yunes fit performs the best out of the three fits. This fit falls within its reported precision when compared with unified as well as nonunified EoSs, for soft and stiff EoSs and modern or older (H3 and H4) EoSs.

\section{Conclusion}
We have established piecewise polytropic fits based on unified tables of equations of state for 52 different nuclear models: 15 nucleonic relativistic mean-field models, seven hyperonic relativistic mean-field models, five relativistic mean-field hybrid models, 24 nucleonic Skyrme models, and one \textit{ab-initio} model. 

The set of 52 equations of state chosen in this paper is confronted with both astrophysical observations and nuclear matter laboratory constraints. Microscopic constraints on nuclear models are discussed, particularly how the quantities $J$ and $L$, respectively the symmetry energy and its slope calculated at saturation density, are constrained by laboratory measurements. The Direct Urca process, which is a very efficient cooling reaction in the core, occurs for 47 models of our set of equations of state. 

We show that establishing piecewise polytropic fits based on nonunified equations of state results in errors on the modeling of macroscopic parameters that can be as high as approximately 15\% ; note that this value corresponds to errors on the tidal deformability and is given excluding configurations close to maximum mass, because the relative uncertainty at fixed mass is then artificially high, as discussed in Sec.~\ref{sec:ppfperform}. This artificial uncertainty is brought forth by the incompatibility of the crust and the core with regards to the microscopic quantities $L$ and $J$.

We present parameters for piecewise polytropic fits based on unified equations of state. 14 parameters per model are required to establish the relation between the pressure, the energy density and the baryonic density; a \textit{Python} routine and tables of fit parameters in ASCII format to compute the tabulated equation of state is provided, see Ancillary files. To establish the fit, a nonlinear least squares method is used, with adjusted transition densities. The precision of our fit was evaluated on macroscopic quantities.

For our 51 models consistent with $M_{\rm max}\gtrsim 2M_\odot$, the fit error on key values of the mass, the radius, and the moment of inertia stays under 0.6\% and 3.5\% in the case of the tidal deformability for a broad range of NS masses except the region very close to the maximum mass, where this inaccuracy can be few times larger.

Finally, we confront the universal relations established by Ref.~\cite{Godzieba2021} between the compactness and the second-order tidal deformability, to that of unified equations of state, as well as two other universal relations \cite{Yagi17,Maselli13}. We conclude that the unified treatment of the crust plays a role in the reported precision of some of those relations. 

\section*{Acknowledgments}

We thank A. Raduta for usefull discussion and the referee for useful comments. The authors acknowledge the financial support of the National Science Center Poland Grant No.~2018/29/B/ST9/02013 (LS, JLZ) and 2017/26/D/ST9/00591 (MF, LS), and of the Funda\c{c}\~ao para a Ci\^encia e Tecnologia  under the Projects UIDP/04564/2020 and UIDB/04564/2020 (C.P.).


\appendix \label{append}

\begin{appendix}

\section{Piecewise polytropic fit parameters} \label{app:param}
We present the parameters of the piecewise polytropic fits based on unified tables of relativistic mean-field models in Table~\ref{tab:RMFparam} and of Skyrme and \textit{ab initio} models in Table~\ref{tab:skyrmeParam}. 

\begin{turnpage}

\begin{table*}[h!]
\caption{Parameters of unified fits by seven polytropes of 15 nucleonic, seven hyperonic and five hybrid relativistic mean-field equations of state. The logarithm of the transition densities $\rho_i$ between the polytropes is given in g/cm$^{3}$. For each polytrope $i$, the adiabatic index $\Gamma_i$ is presented. Only the first constant $\kappa_0$ is presented because all others can be calculated from pressure continuity.}
\label{tab:RMFparam}
\center\begin{tabular}{c|ll|ll|ll|ll|ll|ll|ll}
\hline 
EOS &	 $\log_{10}(\kappa_0)$ &	 $\Gamma_0$ &	$\log_{10}(\rho_1)$ &	 $\Gamma_1$ &	 $\log_{10}(\rho_2)$ &	 $\Gamma_2$ &	 $\log_{10}(\rho_3)$ &	 $\Gamma_3$ &	 $\log_{10}(\rho_4)$ &	 $\Gamma_4$ &	 $\log_{10}(\rho_5)$ &	 $\Gamma_5$ &	 $\log_{10}(\rho_6)$ &	 $\Gamma_6$\\
\hline 
\hline

\multicolumn{15}{c}{Nucleonic RMF EOS}\\
\hline
  BSR2 &	 12.4812 &	 1.6379 &	 6.9304 &	 1.3113 &	 11.3669 &	 0.8349 &	 12.7363 &	 1.3136 &	 14.0413 &	 3.2464 &	 14.8162 &	 2.8221 &	 14.9832 &	 2.3788 \\
  BSR6 &	 12.4804 &	 1.6381 &	 6.9312 &	 1.3109 &	 11.4161 &	 0.7053 &	 12.8819 &	 1.2421 &	 13.5005 &	 2.5053 &	 14.4823 &	 3.1753 &	 14.9091 &	 2.4855\\
  DD2 &	 12.4878 &	 1.6369 &	 6.9309 &	 1.3114 &	 11.3929 &	 0.6260 &	 12.3993 &	 1.2833 &	 13.7322 &	 2.3253 &	 14.3792 &	 3.4041 &	 14.8719 &	 2.6026\\
  DDH$\delta$ &	 12.4849 &	 1.6372 &	 6.9466 &	 1.3092 &	 11.4523 &	 0.5441 &	 12.1843 &	 1.1022 &	 14.1019 &	 4.1828 &	 14.5019 &	 3.1328 &	 14.8150 &	 2.4599\\
  DDME2 &	 12.4955 &	 1.6353 &	 6.9470 &	 1.3106 &	 11.4015 &	 0.6160 &	 12.3748 &	 1.2921 &	 13.6025 &	 2.0223 &	 14.3561 &	 3.5998 &	 14.8460 &	 2.6395\\
  FSU2 &	 12.5074 &	 1.6330 &	 6.9793 &	 1.3076 &	 11.4658 &	 0.6605 &	 12.7237 &	 0.8687 &	 13.5102 &	 2.9854 &	 14.1278 &	 2.6376 &	 14.9194 &	 1.9831\\
  FSU2H &	 12.4979 &	 1.6349 &	 6.9546 &	 1.3097 &	 11.4067 &	 0.7657 &	 12.4968 &	 1.3578 &	 14.2427 &	 3.9780 &	 14.6581 &	 3.1615 &	 14.8787 &	 2.1387\\
  FSU2R &	 12.4986 &	 1.6347 &	 6.9527 &	 1.3103 &	 11.3870 &	 0.7898 &	 12.4679 &	 1.3331 &	 14.2033 &	 3.7040 &	 14.6178 &	 2.8757 &	 14.8944 &	 2.0137\\
  GM1 &	 12.4928 &	 1.6356 &	 6.9626 &	 1.3082 &	 11.4783 &	 0.5103 &	 12.2341 &	 0.9431 &	 13.6981 &	 3.2095 &	 14.3853 &	 2.8973 &	 14.9312 &	 2.5144\\
  NL3 &	 12.4945 &	 1.6355 &	 6.9470 &	 1.3103 &	 11.4119 &	 0.6234 &	 12.3397 &	 0.9161 &	 13.5283 &	 2.8788 &	 14.5470 &	 3.4771 &	 14.8390 &	 2.5896\\
  NL3${\omega\rho}$ &	 12.4740 &	 1.6396 &	 6.8920 &	 1.3155 &	 11.2354 &	 0.7958 &	 12.8470 &	 1.6250 &	 14.2557 &	 3.9080 &	 14.7642 &	 3.1231 &	 14.9024 &	 2.5096\\
  TM1 &	 12.4922 &	 1.6360 &	 6.9387 &	 1.3113 &	 11.3769 &	 0.5885 &	 12.2818 &	 1.0673 &	 13.6098 &	 2.8867 &	 14.2938 &	 2.6964 &	 14.8874 &	 2.0656\\
  TM1${\omega\rho}$ &	 12.4834 &	 1.6377 &	 6.9197 &	 1.3125 &	 11.3283 &	 0.8353 &	 13.0023 &	 1.7447 &	 14.2658 &	 3.2911 &	 14.7090 &	 2.6657 &	 14.9376 &	 2.0072\\
  TM2 &	 12.4986 &	 1.6347 &	 6.9558 &	 1.3096 &	 11.4258 &	 0.7248 &	 12.7689 &	 1.0601 &	 13.5766 &	 2.8071 &	 14.8360 &	 2.4069 &	 14.9871 &	 1.9881\\
  TM2${\omega\rho}$ &	 12.4809 &	 1.6382 &	 6.9119 &	 1.3133 &	 11.3180 &	 0.8364 &	 13.0174 &	 1.7590 &	 14.2803 &	 3.3754 &	 14.7323 &	 2.7264 &	 14.9386 &	 2.0438\\
\hline 
\multicolumn{15}{c}{Hyperonic RMF EOS}\\
\hline 
    DD2 &	 12.4849 &	 1.6373 &	 6.9355 &	 1.3108 &	 11.4036 &	 0.6167 &	 12.3954 &	 1.2856 &	 13.7387 &	 2.3656 &	 14.4082 &	 3.4499 &	 14.7460 &	 2.1317\\
    DDME2 &	 12.4797 &	 1.6383 &	 6.9258 &	 1.3112 &	 11.3963 &	 0.6274 &	 12.4257 &	 1.3473 &	 13.7718 &	 2.1575 &	 14.3628 &	 3.6315 &	 14.7501 &	 2.1179\\
     FSU2H &	 12.4855 &	 1.6371 &	 6.9377 &	 1.3105 &	 11.3993 &	 0.7711 &	 12.4958 &	 1.3600 &	 14.2574 &	 4.1927 &	 14.5282 &	 3.6776 &	 14.7324 &	 1.9163\\
    H3 &	 12.7365 &	 1.5950 &	 7.1558 &	 1.3021 &	 11.5194 &	 0.4741 &	 12.2298 &	 0.9455 &	 13.7026 &	 3.2473 &	 14.3214 &	 2.9180 &	 14.6654 &	 1.9421\\
    H4 &	 12.7332 &	 1.5958 &	 7.1362 &	 1.3035 &	 11.5018 &	 0.4987 &	 12.2443 &	 0.9454 &	 13.7026 &	 3.2456 &	 14.3267 &	 2.9158 &	 14.7047 &	 2.1990\\    
    NL3 &	 12.4804 &	 1.6382 &	 6.9277 &	 1.3111 &	 11.4092 &	 0.6241 &	 12.3368 &	 0.9139 &	 13.5225 &	 2.8704 &	 14.5487 &	 3.4335 &	 14.6612 &	 2.1934\\
    NL3${\omega\rho}$ &	 12.4666 &	 1.6409 &	 6.8926 &	 1.3141 &	 11.3219 &	 0.7170 &	 12.5349 &	 1.3253 &	 13.5939 &	 2.0372 &	 14.3365 &	 3.8767 &	 14.7107 &	 2.1491\\
\hline
\multicolumn{15}{c}{Hybrid EOS}\\
\hline
DD2-B15-40-20 & 12.8916 & 1.5682 & 7.3053 & 1.3013 & 11.4524 & 0.5827 & 12.3650 & 1.2772 & 13.7478 & 2.3416 & 14.3675 & 3.3902 & 14.8872 & 1.2831 \\
NL3$\omega\rho$-B20-50-0 &  12.6482 & 1.6090 & 7.0411 & 1.3126 & 11.1942 & 0.8373 & 13.0106 & 1.7766 & 14.2898 & 3.7905 & 14.6848 & 2.1843 & 15.0271 & 1.4575 \\
NL3$\omega \rho$-B28-75-0 &  12.6539 & 1.6079 & 7.0516 & 1.3122 & 11.1957 & 0.8375 & 13.0148 & 1.7533 & 14.2720 & 3.8022 & 14.7582 & 1.8590 & 15.0730 & 1.4889 \\
NL3$\omega \rho$-B0-50-0 &  12.6539 & 1.6079 & 7.0516 & 1.3122 & 11.1957 & 0.8375 & 13.0158 & 1.7541 & 14.2723 & 3.8042 & 14.8040 & 0.0344 & 14.9109 & 2.1239 \\
NL3$\omega \rho$-B0-50-50 &  12.6539 & 1.6079 & 7.0557 & 1.3113 & 11.2603 & 0.7904 & 12.8250 & 1.5964 & 14.1493 & 3.0321 & 14.4147 & 3.9648 & 14.7511 & 3.1637 \\
\hline
\hline

\end{tabular} 
\end{table*}
\end{turnpage}

\begin{turnpage}

\begin{table*}[h!]
\caption{Parameters of unified fits by seven polytropes of 24 nucleonic Skyrme and one \textit{ab-initio} equations of state. The logarithm of the transition densities $\rho_i$ between the polytropes is given in g/cm$^{3}$. For each polytrope $i$, the adiabatic index $\Gamma_i$ is presented. Only the first constant $\kappa_0$ is presented because all others can be calculated from pressure continuity.}
\label{tab:skyrmeParam}
\center
{\small
\begin{tabular}{c|ll|ll|ll|ll|ll|ll|ll}
\hline 
EOS &	 $\log_{10}(\kappa_0)$ &	 $\Gamma_0$ &	$\log_{10}(\rho_1)$ &	 $\Gamma_1$ &	 $\log_{10}(\rho_2)$ &	 $\Gamma_2$ &	 $\log_{10}(\rho_3)$ &	 $\Gamma_3$ &	 $\log_{10}(\rho_4)$ &	 $\Gamma_4$ &	 $\log_{10}(\rho_5)$ &	 $\Gamma_5$ &	 $\log_{10}(\rho_6)$ &	 $\Gamma_6$\\
\hline
\hline 

\multicolumn{15}{c}{(Nucleonic) Skyrme EOS}\\
\hline
    BSk20 &	 12.4732 &	 1.6396 &	 6.9219 &	 1.3117 &	 11.3469 &	 0.7499 &	 12.4636 &	 1.3408 &	 14.1522 &	 2.8323 &	 14.4311 &	 3.2096 &	 14.8995 &	 3.0780\\
    BSk21 &	 12.4958 &	 1.6357 &	 6.9433 &	 1.3107 &	 11.3651 &	 0.7452 &	 12.3329 &	 1.2571 &	 14.1610 &	 3.4841 &	 14.6921 &	 3.1032 &	 14.9021 &	 2.8012\\
    BSk22 &	 12.5847 &	 1.6208 &	 7.0094 &	 1.3087 &	 11.3556 &	 0.7443 &	 12.5103 &	 1.3024 &	 14.0180 &	 3.1330 &	 14.6885 &	 2.9089 &	 14.8925 &	 2.7427\\
    BSk23 &	 12.5847 &	 1.6208 &	 7.0094 &	 1.3087 &	 11.3556 &	 0.7443 &	 12.5103 &	 1.3024 &	 14.0180 &	 3.1330 &	 14.6885 &	 2.9089 &	 14.8925 &	 2.7427\\
    BSk24 &	 12.5798 &	 1.6215 &	 7.0054 &	 1.3093 &	 11.3762 &	 0.7402 &	 12.3322 &	 1.2579 &	 14.1588 &	 3.4628 &	 14.7075 &	 3.0922 &	 14.9230 &	 2.7773\\
    BSk25 &	 12.5907 &	 1.6197 &	 7.0119 &	 1.3090 &	 11.3885 &	 0.7444 &	 12.2107 &	 1.2034 &	 14.2131 &	 3.7548 &	 14.6893 &	 3.1507 &	 14.9108 &	 2.6403\\
    BSk26 &	 12.4353 &	 1.6458 &	 6.9024 &	 1.3126 &	 11.3405 &	 0.7526 &	 12.4679 &	 1.3404 &	 14.1348 &	 2.7472 &	 14.4199 &	 3.2064 &	 14.9161 &	 3.0628\\
    DH &	 12.7007 &	 1.6021 &	 7.0898 &	 1.3030 &	 11.5622 &	 0.6165 &	 12.4163 &	 1.3397 &	 14.0053 &	 2.1052 &	 14.2804 &	 3.0053 &	 14.9602 &	 2.8605\\
    KDE0v1 &	 14.7161 &	 1.3184 &	 10.1496 &	 1.2477 &	 11.5395 &	 0.6476 &	 12.4235 &	 1.3753 &	 14.0090 &	 2.4045 &	 14.4262 &	 2.8665 &	 15.0278 &	 2.7822\\
    Rs &	 14.7794 &	 1.3089 &	 10.2552 &	 1.2161 &	 11.7133 &	 0.5642 &	 13.0311 &	 0.3835 &	 13.3745 &	 1.4335 &	 13.5407 &	 3.1815 &	 14.2645 &	 2.6712\\
    Sk255 &	 14.7118 &	 1.3176 &	 10.1273 &	 1.2456 &	 11.5501 &	 0.5897 &	 12.5538 &	 1.2295 &	 13.5409 &	 2.4784 &	 14.4910 &	 2.7236 &	 15.1723 &	 2.6880 \\
    Sk272 &	 14.7050 &	 1.3188 &	 10.1052 &	 1.2497 &	 11.5150 &	 0.6131 &	 12.4888 &	 1.2939 &	 13.6393 &	 2.4588 &	 14.4272 &	 2.8096 &	 15.1060 &	 2.7603\\
    Ska &	 14.7299 &	 1.3149 &	 10.1492 &	 1.2381 &	 11.5816 &	 0.5908 &	 12.4222 &	 1.1598 &	 13.5067 &	 2.0177 &	 14.0436 &	 2.8420 &	 15.0849 &	 2.7774\\
    Skb &	 14.7293 &	 1.3142 &	 10.1099 &	 1.2372 &	 11.5344 &	 0.7113 &	 13.2162 &	 0.3365 &	 13.7754 &	 4.0702 &	 14.2652 &	 3.0945 &	 14.7695 &	 2.8537\\
    SkI2 &	 14.7376 &	 1.3144 &	 10.1308 &	 1.2373 &	 11.5685 &	 0.6265 &	 13.4248 &	 1.7804 &	 13.6041 &	 3.2146 &	 14.3183 &	 2.6160 &	 15.0811 &	 2.6441\\
    SkI3 &	 14.7239 &	 1.3164 &	 10.1258 &	 1.2435 &	 11.5830 &	 0.5858 &	 12.3665 &	 1.1000 &	 13.7485 &	 2.9839 &	 14.4126 &	 2.8078 &	 14.6677 &	 2.6923\\
    SkI4 &	 14.7263 &	 1.3167 &	 10.1323 &	 1.2426 &	 11.5727 &	 0.5761 &	 12.3014 &	 1.1311 &	 13.9299 &	 3.1012 &	 14.6916 &	 2.9305 &	 14.9658 &	 2.7467\\
    SkI5 &	 14.7427 &	 1.3139 &	 10.2018 &	 1.2345 &	 11.6527 &	 0.4384 &	 12.0220 &	 0.6561 &	 13.5518 &	 3.3975 &	 14.2875 &	 2.5666 &	 14.9267 &	 2.6828 \\
    SkI6 &	 14.7290 &	 1.3163 &	 10.1451 &	 1.2418 &	 11.5705 &	 0.5986 &	 12.3258 &	 1.1753 &	 13.9557 &	 3.0843 &	 14.7293 &	 2.9186 &	 14.9817 &	 2.7458\\
    SkMP &	 14.7605 &	 1.3113 &	 10.1929 &	 1.2251 &	 11.6484 &	 0.5811 &	 12.6209 &	 1.0106 &	 13.6166 &	 2.7978 &	 14.4767 &	 2.7814 &	 14.9768 &	 2.7302\\
    SkOp &	 14.7348 &	 1.3160 &	 10.1478 &	 1.2410 &	 11.5968 &	 0.5485 &	 12.4485 &	 1.0966 &	 13.3943 &	 1.8574 &	 13.9326 &	 2.6883 &	 15.1036 &	 2.6213\\
    SLy230a &	 14.7200 &	 1.3174 &	 10.1478 &	 1.2437 &	 11.5337 &	 0.6262 &	 12.2329 &	 1.2824 &	 14.1720 &	 3.1458 &	 14.8556 &	 2.9664 &	 15.0878 &	 2.7300\\
    SLy2 &	 14.7218 &	 1.3170 &	 10.1456 &	 1.2429 &	 11.5318 &	 0.6369 &	 12.3351 &	 1.3217 &	 14.0238 &	 2.4088 &	 14.3263 &	 2.9840 &	 14.9738 &	 2.8379\\
    SLy9 &	 14.7253 &	 1.3165 &	 10.1345 &	 1.2418 &	 11.5328 &	 0.6416 &	 12.3443 &	 1.3051 &	 13.9715 &	 2.5671 &	 14.2802 &	 2.9772 &	 14.9302 &	 2.7763\\
\hline 

\multicolumn{15}{c}{\textit{ab-initio} EOS}\\
\hline
  BCPM &	 12.4703 &	 1.6383 &	 6.9467 &	 1.3136 &	 11.3401 &	 0.7181 &	 12.4647 &	 1.3333 &	 14.0080 &	 2.7194 &	 14.0053 &	 2.9133 &	 14.9915 &	 2.6914\\
\hline
\hline

\end{tabular} 
}
\end{table*}
\end{turnpage}


\section{Accuracy on macroscopic parameters} 
We present the relative difference for astrophysical quantities of interest between the unified tables and our piecewise polytropic fits for relativistic mean-field models in Table~\ref{tab:rmfmacro} and for Skyrme and \textit{ab initio} models in Table~\ref{tab:Skyrmemacro}.
\begin{turnpage}
\begin{table*}[h!]
\caption{Key macroscopic quantities calculated from the unified table of 15 nucleonic, seven hyperonic and five hybrid relativistic mean-field models, and the relative errors $\Delta$ and $\delta$ (see Sec.~\ref{sec:ppfperform})} in pourcent related to unified piecewise polytropic fit. The maximum mass (in solar mass) $M_{\rm max}$, the density (in fm$^{-3}$) at maximum mass $n_{\rm max}$, the radius (in km) for a 1\;M$_{\odot}$ NS $R_{1.0}$, the radius for a 1.4\;M$_{\odot}$ NS $R_{1.4}$, the radius at maximum mass $R_{M_{\rm max}}$, the moment of inertia (in $10^{45}$g.cm$^2$) for a 1.338\;M$_{\odot}$ NS $I_{1.338}$ as measured in the double pulsar PSR J0737$-$3039, the moment of inertia at maximum mass $I_{M_{\rm max}}$, the tidal deformability for a 1.4\;M$_{\odot}$ NS $\Lambda_{1.4}$, and the tidal deformability at maximum mass $\Lambda_{M_{\rm max}}$ are presented. In {\color{red} red}, we indicate the equations of state that gives the largest relative fit error in each category. The maximum mass of EoS H3 is indicated in \textcolor{blue}{blue}, to emphasize that is not consistent with J$1614-2230$ mass measurement; results are shown only because this model is used in Sec.~III.B.
\label{tab:rmfmacro}
\center\begin{tabular}{c|ll|ll|ll|ll|ll|ll|ll|ll|ll}
\hline 
&  $M_{\rm max}$ & $\Delta$ & $n_{\rm max}$ & $\delta$ & $R_{1.0}$ & $\Delta$ &  $R_{1.4}$ & $\Delta$ &  $R_{M_{\rm max}}$& $\delta$  & $I_{1.338}$& $\Delta$ & $I_{M_{\rm max}}$& $\delta$  & $\Lambda_{1.4}$ & $\Delta$ & $\Lambda_{M_{\rm max}}$ & $\delta$ \\
\hline
\hline 
\multicolumn{19}{c}{Nucleonic RMF EoS}\\
\hline
BSR2 & 2.383 & -0.14 & 0.852 & -0.25 & 13.30 & -0.16 & 13.40 & -0.19 & 11.96 & -0.11 & 1.634 & -0.28 & 3.149 & -0.36 & 761.70 & -1.00 & 6.72 & 0.45\\
BSR6 & 2.430 & -0.21 & 0.827 & 0.00 & 13.77 & 0.12 & 13.73 & 0.06 & 12.13 & -0.24 & 1.677 & 0.02 & 3.306 & -0.77 & 836.16 & 0.01 & 6.28 & -0.99\\
DD2 & 2.417 & -0.20 & 0.851 & {\color{red} 0.63} & 12.99 & 0.22 & 13.16 & -0.01 & 11.87 & -0.47 & 1.593 & -0.30 & 3.216 & -1.14 & 697.89 & -1.13 & 5.74 & {\color{red} -3.81}\\
DDH$\delta$ & 2.138 & -0.30 & 1.000 & -0.00 & 12.40 & -0.04 & 12.61 & -0.04 & 11.14 & {\color{red}-0.52} & 1.533 & -0.37 & 2.376 & {\color{red}-1.42} & 589.00 & 0.41 & 9.23 & -2.72\\
DDME2 & 2.481 & -0.24 & 0.817 & -0.23 & 12.98 & 0.16 & 13.20 & -0.07 & 12.06 & -0.32 & 1.604 & {\color{red} -0.41} & 3.456 & -0.95 & 719.61 & -1.56 & 5.39 & -1.77\\
FSU2 & 2.071 & -0.22 & 0.904 & -0.00 & 14.18 & -0.17 & 13.93 & -0.16 & 12.08 & -0.35 & 1.716 & -0.10 & 2.441 & -0.90 & 886.54 & -0.56 & 20.11 & -1.30\\
FSU2H & 2.375 & -0.25 & 0.802 & -0.27 & 13.05 & 0.11 & 13.32 & 0.13 & 12.37 & -0.23 & 1.638 & 0.17 & 3.306 & -0.88 & 752.85 & {\color{red} 3.44} & 9.93 & 0.59\\
FSU2R & 2.047 & -0.30 & 0.943 & -0.28 & 12.89 & 0.32 & 12.98 & 0.02 & 11.66 & -0.30 & 1.552 & -0.25 & 2.333 & -1.16 & 608.63 & 1.61 & 18.35 & 0.03\\
GM1 & 2.361 & -0.14 & 0.864 & 0.00 & 13.64 & -0.08 & 13.72 & -0.12 & 11.92 & -0.20 & 1.729 & -0.27 & 3.063 & -0.52 & 922.33 & -1.02 & 6.82 & -0.82\\
NL3 & 2.773 & -0.16 & 0.669 & 0.00 & 14.52 & -0.32 & 14.61 & -0.18 & 13.29 & -0.23 & 1.898 & -0.16 & 4.744 & -0.57 & 1297.27 & -0.48 & 4.71 & -0.80 \\
NL3${\omega\rho}$ & 2.753 & -0.08 & 0.688 & 0.00 & 13.42 & -0.31 & 13.75 & {\color{red} -0.26} & 13.00 & -0.15 & 1.732 & -0.13 & 4.612 & -0.27 & 953.91 & -0.62 & 4.47 & -0.30\\
TM1 & 2.175 & -0.17 & 0.856 & 0.00 & 14.36 & 0.17 & 14.24 & -0.00 & 12.34 & -0.26 & 1.798 & -0.21 & 2.739 & -0.80 & 1051.06 & -0.73 & 16.70 & -1.52\\
TM1${\omega\rho}$ & 2.118 & {\color{red} -0.36} & 0.908 & -0.28 & 13.43 & {\color{red} -0.37} & 13.41 & -0.22 & 11.91 & -0.34 & 1.607 & -0.21 & 2.522 & -1.08 & 712.90 & -0.82 & 16.20 & -0.01\\
TM2 & 2.270 & -0.14 & 0.823 & -0.00 & 14.44 & -0.19 & 14.34 & -0.12 & 12.50 & -0.14 & 1.813 & -0.22 & 3.010 & -0.44 & 1087.63 & -0.67 & 13.45 & 0.05\\
TM2${\omega\rho}$ & 2.220 & -0.34 & 0.869 & 0.00 & 13.43 & {\color{red}-0.37} & 13.47 & -0.24 & 12.08 & -0.37 & 1.626 & -0.32 & 2.803 & -1.09 & 748.94 & -0.98 & 12.99 & -0.46\\
\hline 
\multicolumn{19}{c}{Hyperonic RMF EoS}\\
\hline 
DD2 & 1.996 & 0.02 & 1.007 & {\color{red}-1.48} & 12.99 & 0.37 & 13.15 & 0.15 & 11.38 & 0.72 & 1.592 & -0.11 & 2.125 & 1.18 & 694.86 & -0.50 & 17.14 & {\color{red}6.50}\\
DDME2 & 2.064 & 0.06 & 0.947 & -1.20 & 12.98 & -0.11 & 13.20 & -0.15 & 11.65 & 0.45 & 1.604 & -0.19 & 2.342 & 1.00 & 719.19 & -0.66 & 16.70 & 4.76\\
FSU2H & 1.991 & -0.03 & 0.901 & -0.99 & 13.05 & 0.02 & 13.32 & -0.01 & 11.99 & 0.33 & 1.638 & -0.14 & 2.310 & 0.60 & 752.79 & {\color{red}2.57} & 27.96 & 5.22\\
H3 & {\color{blue} 1.787} & {\color{red} -0.54} & 0.993 & -0.28 & 13.66 & -0.04 & 13.61 & 0.04 & 11.75 & {\color{red} -0.76} & 1.707 & -0.03 & 1.839 & {\color{red} -2.39} & 852.60 & 0.07 & 47.43 & -3.58\\
H4 & 2.032 & -0.38 & 0.964 & -0.00 & 13.66 & -0.06 & 13.72 & -0.08 & 11.71 & -0.53 & 1.730 & {\color{red}-0.27} & 2.268 & -1.57 & 920.95 & -0.94 & 18.32 & -2.48\\
NL3 & 2.232 & 0.02 & 0.737 & 0.00 & 14.52 & -0.15 & 14.61 & -0.10 & 12.90 & 0.34 & 1.898 & -0.13 & 3.058 & 0.72 & 1297.07 & -0.45 & 20.05 & 3.65\\
NL3${\omega\rho}$ & 2.277 & -0.08 & 0.751 & 0.00 & 13.42 & {\color{red} 0.50} & 13.75 & {\color{red}0.23} & 12.69 & 0.23 & 1.732 & -0.05 & 3.158 & 0.17 & 953.84 & -0.34 & 16.87 & 2.30\\
\hline 
\multicolumn{19}{c}{Hybrid EOS}\\
\hline 
DD2-B15-40-20 & 2.153 & 0.04 & 0.771 & {\color{red}-0.40} & 12.99 & 0.05 & 13.16 & -0.14 & 12.65 & 0.02 & 1.593 & {\color{red}-0.32} & 2.933 & 0.17 & 698.15 & -1.35 & 28.25 & 0.13\\
NL3$\omega\rho$B20-50-0 & 2.151 & -0.29 & 0.812 & -0.37 & 13.40 & {\color{red}0.59} & 13.73 & 0.16 & 12.58 & 0.04 & 1.730 & {\color{red}-0.32} & 2.836 & -0.61 & 950.32 & {\color{red}-1.54} & 24.32 & 1.39\\
NL3$\omega \rho$-B28-75-0 & 2.326 & -0.23 & 0.729 & -0.36 & 13.40 & 0.41 & 13.73 & 0.16 & 13.07 & -0.02 & 1.730 & 0.09 & 3.450 & -0.45 & 950.32 & 0.02 & 19.36 & {\color{red}1.40}\\
NL3$\omega \rho$-B0-50-0 & 2.241 & {\color{red}-0.45} & 0.666 & -0.23 & 13.40 & 0.41 & 13.73 & {\color{red}0.17} & 13.44 & {\color{red}-0.15} & 1.730 & 0.11 & 3.428 & {\color{red}-1.10} & 950.28 & 0.09 & 34.19 & 1.32\\
NL3$\omega \rho$-B0-50-50 & 2.455 & -0.17 & 0.443 & 0.00 & 13.40 & -0.03 & 13.73 & -0.07 & 13.96 & -0.06 & 1.730 & -0.10 & 4.297 & -0.35 & 950.29 & -0.45 & 25.97 & 0.79\\
\hline 
\hline
\end{tabular} 
\end{table*}

\begin{table*}[h!]
\caption{Key macroscopic quantities calculated from the unified table of 24 Skyrme and one \textit{ab-initio} models, and the relative errors $\Delta$ and $\delta$ (see Section~\ref{sec:ppfperform})} in pourcent related to unified piecewise polytropic fit. The maximum mass (in solar mass) $M_{\rm max}$, the density (in fm$^{-3}$) at maximum mass $n_{\rm max}$, the radius (in km) for a \;M$_{\odot}$ NS $R_{1.0}$, the radius for a 1.4\;M$_{\odot}$ NS $R_{1.4}$, the radius at maximum mass $R_{M_{\rm max}}$, the moment of inertia (in $10^{45}$g.cm$^2$) for a 1.338\;M$_{\odot}$ NS $I_{1.338}$ as measured in the double pulsar PSR J0737$-$3039, the moment of inertia at maximum mass $I_{M_{\rm max}}$, the tidal deformability for a 1.4\;M$_{\odot}$ NS $\Lambda_{1.4}$, and the tidal deformability at maximum mass $\Lambda_{M_{\rm max}}$ are presented. In {\color{red} red}, we indicate the equations of state that gives the largest relative fit error in each category. 
\label{tab:Skyrmemacro}
\center\begin{tabular}{c|ll|ll|ll|ll|ll|ll|ll|ll|ll}
\hline 
&  $M_{\rm max}$ & $\Delta$ & $n_{\rm max}$ & $\delta$ & $R_{1.0}$ & $\Delta$ &  $R_{1.4}$ & $\Delta$ &  $R_{M_{\rm max}}$& $\delta$  & $I_{1.338}$& $\Delta$ & $I_{M_{\rm max}}$& $\delta$  & $\Lambda_{1.4}$ & $\Delta$ & $\Lambda_{M_{\rm max}}$ & $\delta$ \\
\hline
\hline 
\multicolumn{19}{c}{(Nucleonic) Skyrme EOS}\\
\hline
BSk20 & 2.164 & 0.04 & 1.126 & 0.52 & 11.76 & 0.00 & 11.74 & -0.00 & 10.17 & -0.25 & 1.308 & -0.06 & 2.176 & -0.26 & 328.30 & -0.19 & 3.53 & -3.18\\
BSk21 & 2.274 & 0.06 & 0.975 & {\color{red} 0.77} & 12.47 & 0.04 & 12.59 & -0.02 & 11.04 & -0.31 & 1.484 & -0.06 & 2.622 & -0.34 & 533.99 & -0.34 & 4.90 & -3.93\\
BSk22 & 2.265 & -0.12 & 0.969 & 0.24 & 13.03 & -0.08 & 13.05 & -0.08 & 11.19 & -0.19 & 1.564 & -0.13 & 2.622 & -0.47 & 642.77 & -0.52 & 5.39 & -0.92\\
BSk23 & 2.265 & -0.12 & 0.969 & 0.24 & 13.03 & -0.08 & 13.05 & -0.08 & 11.19 & -0.19 & 1.564 & -0.13 & 2.622 & -0.47 & 642.77 & -0.52 & 5.39 & -0.92\\
BSk24 & 2.279 & -0.15 & 0.978 & 0.24 & 12.47 & -0.06 & 12.59 & -0.07 & 11.05 & -0.23 & 1.483 & -0.11 & 2.637 & -0.61 & 532.32 & -0.53 & 4.84 & -1.21\\
BSk25 & 2.225 & {\color{red} -0.17} & 0.998 & 0.46 & 12.22 & 0.02 & 12.39 & -0.07 & 10.99 & -0.34 & 1.454 & -0.16 & 2.516 & -0.84 & 495.04 & -0.74 & 5.86 & -2.33\\
BSk26 & 2.169 & -0.09 & 1.124 & 0.36 & 11.79 & -0.05 & 11.78 & -0.06 & 10.20 & -0.25 & 1.314 & -0.17 & 2.191 & -0.51 & 333.57 & -0.59 & 3.53 & -2.01\\
DH & 2.049 & -0.04 & 1.207 & 0.00 & 11.90 & -0.06 & 11.73 & -0.03 & 9.99 & -0.12 & 1.287 & -0.08 & 1.904 & -0.25 & 304.98 & -0.20 & 4.64 & -1.01\\
KDE0v1 & 1.969 & -0.13 & 1.279 & 0.18 & 11.90 & -0.07 & 11.61 & -0.13 & 9.79 & -0.18 & 1.255 & -0.25 & 1.714 & -0.47 & 274.01 & -1.01 & 5.21 & -0.72\\
Rs & 2.116 & -0.12 & 1.074 & 0.15 & 13.05 & -0.09 & 12.91 & -0.09 & 10.75 & -0.15 & 1.547 & -0.15 & 2.186 & -0.42 & 605.14 & -0.51 & 6.46 & -0.47\\
Sk255 & 2.144 & -0.15 & 1.057 & 0.19 & 13.42 & 0.11 & 13.12 & 0.08 & 10.84 & -0.11 & 1.542 & -0.01 & 2.248 & -0.48 & 593.99 & 1.11 & 5.93 & 0.29\\
Sk272 & 2.231 & -0.15 & 0.997 & 0.21 & 13.51 & {\color{red} 0.22} & 13.29 & 0.15 & 11.08 & -0.11 & 1.577 & 0.07 & 2.495 & -0.50 & 657.16 & 0.26 & 5.24 & -0.28\\
Ska & 2.208 & -0.09 & 1.025 & 0.18 & 13.01 & -0.21 & 12.89 & -0.13 & 10.88 & -0.17 & 1.522 & 0.05 & 2.409 & -0.34 & 569.18 & 1.65 & 5.05 & 0.02\\
Skb & 2.188 & -0.13 & 1.060 & 0.69 & 12.05 & -0.03 & 12.19 & 0.07 & 10.60 & {\color{red} -0.49} & 1.449 & 0.17 & 2.333 & {\color{red} -0.95} & 481.85 & 0.67 & 4.86 & {\color{red} -4.29}\\
SkI2 & 2.162 & -0.07 & 1.015 & 0.01 & 13.58 & -0.08 & 13.46 & -0.15 & 11.11 & -0.06 & 1.662 & {\color{red} -0.31} & 2.354 & -0.19 & 786.60 & -1.07 & 6.95 & 0.20\\
SkI3 & 2.239 & -0.10 & 0.967 & 0.08 & 13.59 & 0.00 & 13.53 & -0.02 & 11.30 & -0.08 & 1.666 & -0.12 & 2.574 & -0.30 & 801.58 & -0.44 & 6.14 & -0.04\\
SkI4 & 2.169 & -0.13 & 1.061 & 0.24 & 12.31 & -0.00 & 12.35 & -0.07 & 10.66 & -0.21 & 1.447 & -0.16 & 2.297 & -0.55 & 463.67 & {\color{red} 2.77} & 5.22 & 0.19\\
SkI5 & 2.240 & -0.08 & 0.953 & -0.05 & 14.16 & -0.04 & 14.05 & -0.10 & 11.46 & -0.01 & 1.793 & -0.25 & 2.598 & -0.15 & 1029.71 & -0.83 & 6.62 & 0.74\\
SkI6 & 2.189 & -0.12 & 1.044 & 0.14 & 12.44 & -0.06 & 12.47 & -0.07 & 10.75 & -0.17 & 1.464 & -0.13 & 2.359 & -0.47 & 501.68 & -0.57 & 5.22 & -0.70\\
SkMP & 2.107 & -0.11 & 1.107 & 0.15 & 12.58 & 0.02 & 12.48 & -0.05 & 10.52 & -0.13 & 1.459 & -0.20 & 2.123 & -0.40 & 489.63 & -0.72 & 5.69 & -0.49\\
SkOp & 1.972 & -0.13 & 1.224 & 0.19 & 12.41 & -0.13 & 12.11 & {\color{red} -0.17} & 10.12 & -0.20 & 1.360 & -0.18 & 1.781 & -0.47 & 371.34 & -0.83 & 6.76 & -0.53\\
SLy230a & 2.099 & -0.08 & 1.145 & 0.08 & 11.86 & -0.21 & 11.81 & -0.13 & 10.24 & -0.16 & 1.319 & -0.09 & 2.063 & -0.33 & 338.00 & -0.35 & 4.92 & -0.63\\
SLy2 & 2.053 & -0.11 & 1.197 & 0.26 & 11.91 & -0.13 & 11.76 & -0.14 & 10.04 & -0.25 & 1.301 & -0.23 & 1.924 & -0.52 & 318.13 & -0.82 & 4.77 & -1.62\\
SLy9 & 2.156 & -0.12 & 1.074 & 0.38 & 12.54 & -0.11 & 12.45 & -0.15 & 10.63 & -0.30 & 1.431 & -0.29 & 2.249 & -0.63 & 446.12 & 2.13 & 5.13 & -0.85\\
\hline 
\multicolumn{19}{c}{(Nucleonic) \textit{ab-initio} EoS}\\
\hline 
BCPM & 1.980 & -0.09 & 1.241 & -0.24 & 11.93 & 0.00 & 11.71 & -0.08 & 9.96 & -0.08 & 1.283 & -0.21 & 1.773 & -0.29 & 299.68 & -0.76 & 5.97 & -0.08\\
\hline 
\hline

\end{tabular}  
\end{table*}

\end{turnpage}


\section{Adiabatic index}
\begin{figure*}[h!]
\resizebox{\hsize}{!}{\includegraphics[scale=0.3]{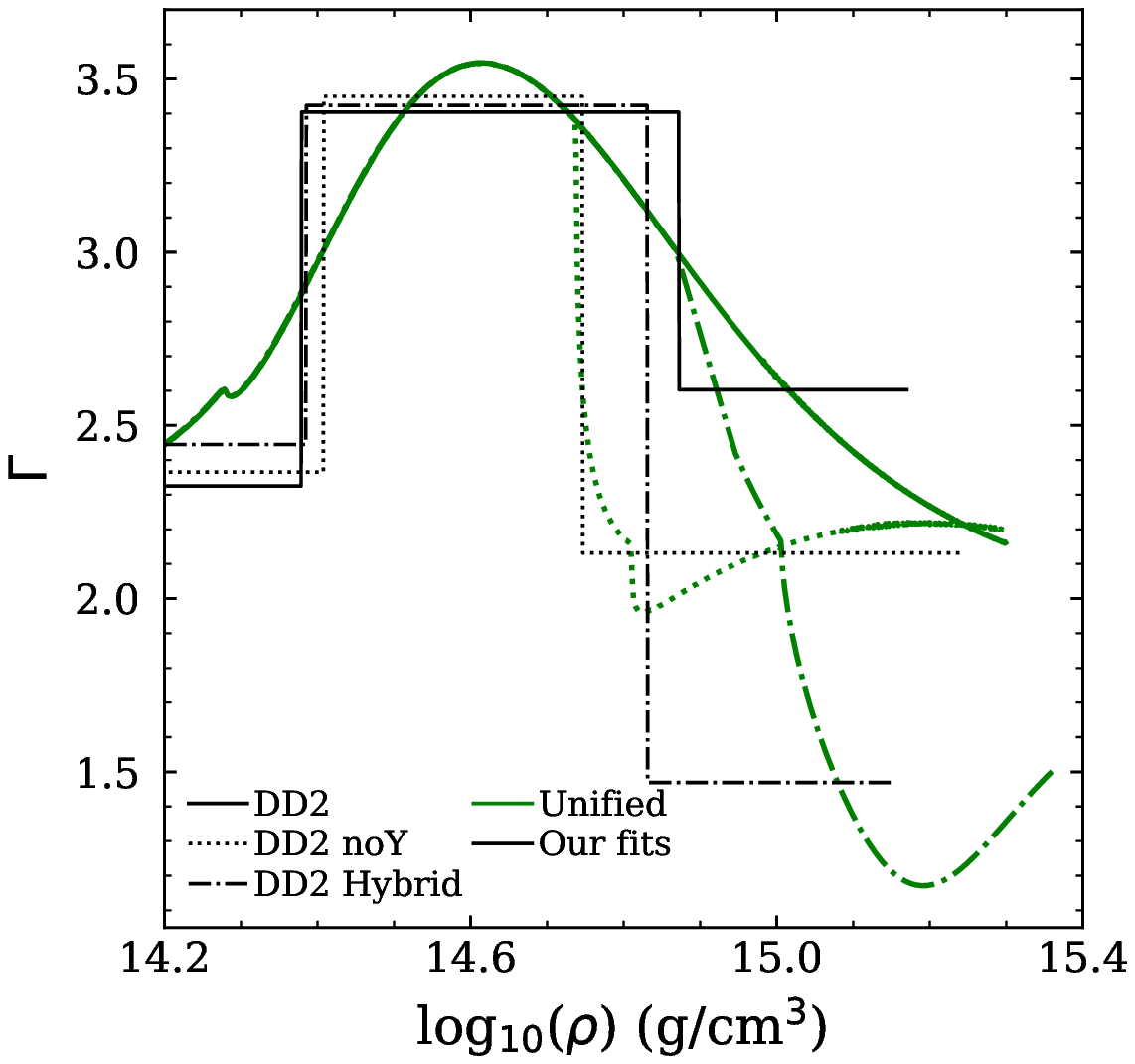}}
\caption{Adiabatic index $\Gamma$ as a function of the baryonic density for the nucleonic, hyperonic, and hybrid EoS DD2. }
\label{fig:gammaAll}
\end{figure*}

We present the adiabatic index as a function of the density in Fig.~\ref{fig:gammaAll} for a few tabulated relativistic mean-fields equations of state based on tabulated unified tables, and our fits. 

\section{Universal relation vs unified equations of state} 

We present results for various fits of so-called universal relations between the tidal deformability and the compactness for different (soft and stiff, old and modern) relativistic mean-field models. 
\begin{figure*}[h!]
\resizebox{\hsize}{!}{\includegraphics[scale=0.3]{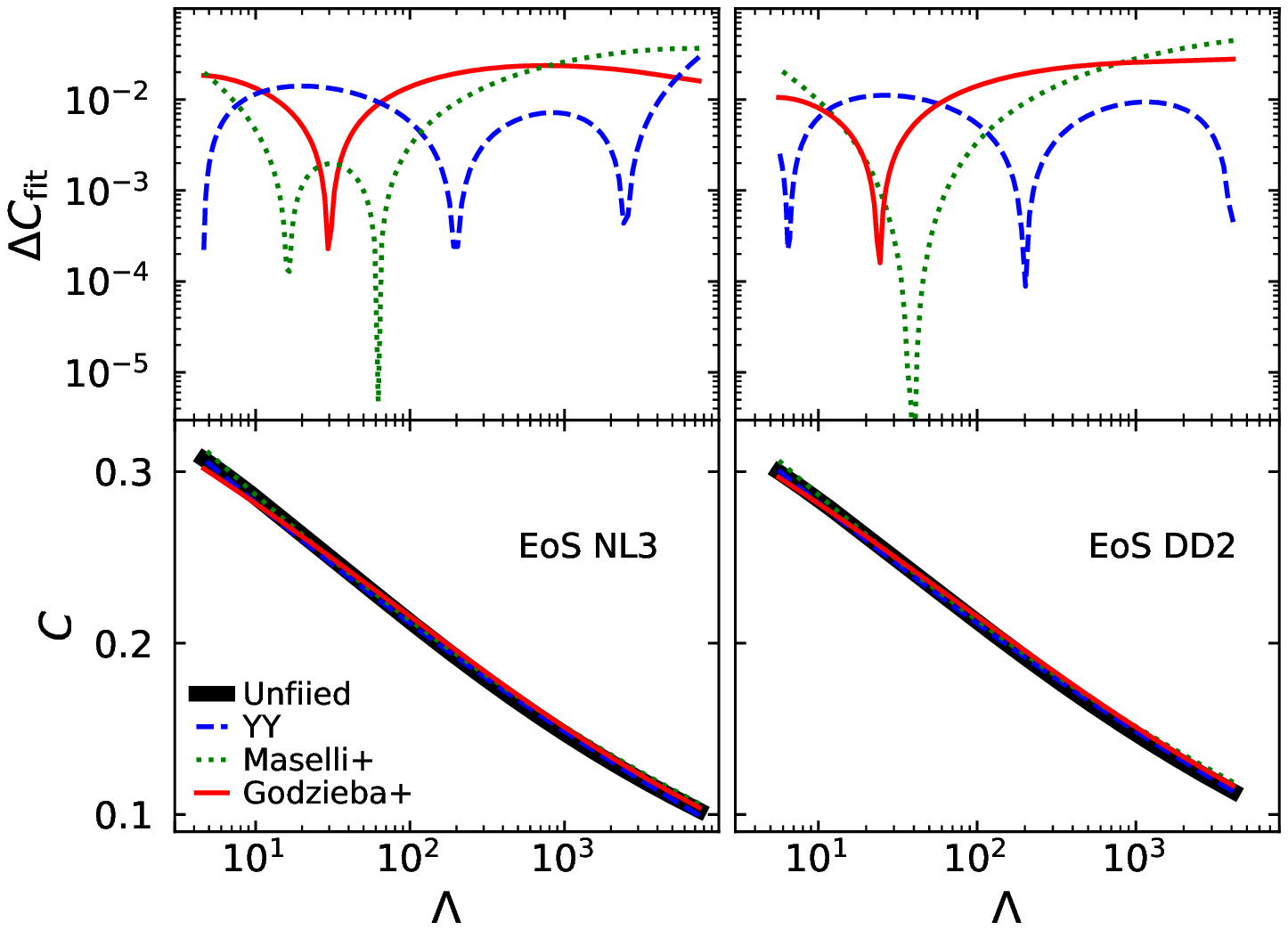}}
\caption{Relative difference $\Delta$C as a function of the tidal deformability $\Lambda$ for EoS NL3 on the left and EoS DD2 on the right.}
\label{fig:Radice}
\end{figure*}
\begin{figure*}[h!]
\resizebox{\hsize}{!}{\includegraphics[scale=0.3]{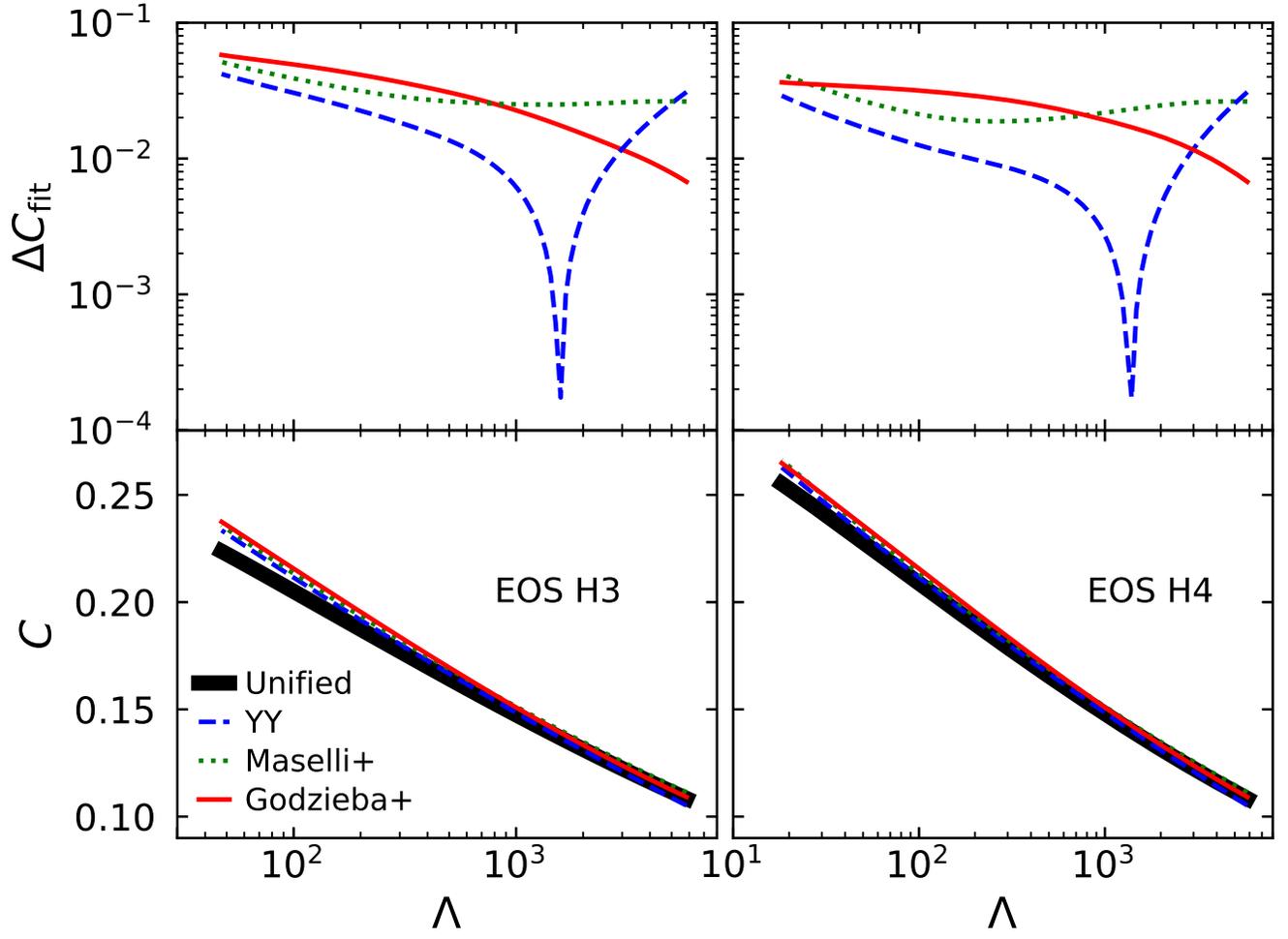}}
\caption{Relative uncertainty $\Delta$C as a function of the tidal deformability $\Lambda$ for EOS H3 on the left and EOS H4 on the right.}
\label{fig:Radice2}
\end{figure*}

\end{appendix}

\newpage


\providecommand{\noopsort}[1]{}\providecommand{\singleletter}[1]{#1}%

\end{document}